\documentclass[12pt,a4paper]{article}
\usepackage[left=2.54cm, right=2.54cm, top=2.54cm, bottom=2.54cm]{geometry}
\usepackage[utf8]{inputenc}
\usepackage{amsmath}
\usepackage{amssymb}
\usepackage{graphicx}
\usepackage{epstopdf}
\usepackage{natbib}
\usepackage{mathtools}
\usepackage{amsthm}
\usepackage{url}
\usepackage{enumitem}
\usepackage{caption}
\usepackage{subcaption}
\usepackage[table]{xcolor}
\usepackage{pdfpages}
\usepackage{authblk}
\usepackage{amsthm}
\usepackage{cleveref}

\usepackage{xcolor}
\usepackage{rotating}
\usepackage{ctable}
\title{A mathematical model for smooth muscle cell phenotype switching in atherosclerotic plaque}
\author[1]{Joseph P.\ Ndenda}
\author[2]{Michael G.\ Watson}
\author[3]{Ashish Misra}
\author[1]{Mary R.\ Myerscough\thanks{mary.myerscough@sydney.edu.au}}
\affil[1]{School of Mathematics and Statistics, University of Sydney, New South Wales 2006, Australia}
\affil[2]{School of Mathematics and Statistics, University of New South Wales, New South Wales 2052, Australia}
\affil[3]{Heart Research Institute, 7 Eliza Street, Newtown, New South Wales 2042, Australia}
\date{ }
\graphicspath{{Figures/}}
\begin{document}
	\maketitle
	\renewcommand{\thefootnote}{\arabic{footnote}}
	\begin{abstract}
		Smooth muscle cells (SMCs) play a fundamental role in the development of atherosclerotic plaques. They ingest lipids in a similar way to monocyte-derived macrophages (MDMs) in the plaque. This can stimulate SMCs to undergo a phenotypic switch to a macrophage-like phenotype. We formulate an ordinary differential equation (ODE) model for the populations of SMCs, MDMs and smooth muscle cell-derived macrophages (SDMs) and the internalised lipid load in each popluation. We use this model to explore the effect on plaque fate of SMC phenotype switching. We find that when SMCs switch to a macrophage-like phenotype, the total lipid contained in the model plaque that is internalised inside cells increases. Additionally, removal of SMCs from the plaque via phenotype switching reduces the number of SMCs in the plaque fibrous cap, increases the lipid in the necrotic core, and increases plaque inflammation. This makes the plaque more vulnerable to rupture, which can lead to heart attacks and strokes. When SDMs are highly proliferative or resistant to cell death, the plaque grows rapidly and becomes highly pathological. The model suggests that plaque dynamics, driven by the switch of SMCs to a macrophage-like phenotype, may drive the development of unstable, vulnerable and pathological plaques.
	\end{abstract}

\newpage
\section{Introduction}
Atherosclerosis remains a major health problem and a leading cause of cardiovascular disease globally \cite{who_cvd_2021}%,herrington2016epidemiology}
. It is caused by chronic inflammation in the walls of large and medium-sized arteries. This inflammation results in the formation of fatty plaques  \cite{gisteraa2017immunology,back2019inflammation}%,libby2002inflammation,falk2006pathogenesis}
. 

Cholesterol-carrying lipoproteins in the blood, mainly low-density lipoproteins (LDLs), enter the artery wall where the endothelium (the cell layer that lines the blood vessel) has become dysfunctional. These LDLs can undergo oxidative and other modifications that render them pro-inflammatory and immunogenic, and cause them to be retained in the vessel wall \cite{libby2002atherosclerosis,%milutinovic2020pathogenesis,
goldberg2022atherogenesis}. Modified LDLs (modLDL) activate resident immune cells in the intima (the part of the artery wall directly beneath the endothelium). These cells respond by secreting pro-inflammatory cytokines, which activate the endothelium and recruit circulating monocytes \cite{back2019inflammation,libby2002atherosclerosis}.

In the intima, monocytes differentiate into macrophages, which express scavenger receptors and internalise modLDL. %The accumulation of lipid inside macrophages gives them a foamy appearance under the microscope. For this reason, these lipid-bearing cells are often referred to as foam cells \cite{libby2002atherosclerosis,ross1999atherosclerosis,yahagi2016pathophysiology}.
Macrophages, in turn, secrete further pro-inflammatory cytokines such as tumor necrosis factor-alpha $(\text{TNF-}\alpha)$ and interleukin-1 (IL-1), %and interleukin-1 (IL-6)), 
which recruit more macrophages and other immune cells into the lesion  \cite{hansson2011immune,zhou1999detection}.
%In atherosclerotic plaques, most leukocytes come from monocyte-derived macrophages (MDMs) that differentiate from circulating blood monocytes \cite{woollard2010monocytes}. 
The number of monocyte-derived macrophages (MDMs) in the plaque is determined by the relative rates of monocyte recruitment \cite{kim2020monocyte}%,bobryshev2016macrophages,shi2011monocyte}
, cell death (apoptosis) \cite{tabas2010macrophage}, proliferation \cite{Lhot16}, and emigration out of the plaque \cite{%llodra2004emigration,
randolph2008emigration}. %In response to lipid ingestion or high lipid loads, MDMs also secrete cytokines such as interleukin 1 (IL-1) and monocyte chemoattractant protein 1 (MCP-1), which increase monocyte recruitment into the intima \cite{back2019inflammation,deshmane2009monocyte,gosling1999mcp,hansson2011immune}.
In addition to modLDL consumption, macrophages can also acquire internalised lipid by consuming apoptotic cells (a process known as efferocytosis) \cite{back2019inflammation,ford2019efferocytosis}. %ford2019lipid,chambers2023new}.
% The LDL phagocytosis introduces lipids into macrophages, whereas efferocytosis recycles and concentrates existing lipids in macrophages \cite{ford2019efferocytosis}. 
Macrophages can reduce their lipid burden by offloading lipid to high-density lipoprotein (HDL) particles \cite{%brown1980cholesteryl,
brown1983lipoprotein}.%, and emigrating macrophages carry their accumulated lipid out of the plaque \cite{llodra2004emigration,randolph2008emigration}.

The medial layer of the artery wall (immediately beneath the intima) contains a population of vascular smooth muscle cells (SMCs). The accumulation of macrophages in the intimal layer triggers the migration of highly proliferative SMCs into the plaque \cite{misra2018integrin}. Recent studies using lineage tracing techniques have shown that SMC populations in plaques are either mono- or oligoclonal, which implies that very few SMCs migrate into the plaque from the media \cite{misra2018integrin}. The proliferative SMCs accumulate beneath the endothelium and form a fibrous cap that covers the lipid-filled plaque core. The number of SMCs in the cap is directly correlated with plaque stability \cite{allahverdian2018smooth,gomez2012smooth}. A thin cap increases the risk of plaque rupture, which can lead to clinical complications such as heart attack or stroke.

Plaque SMCs possess the machinery to undertake phagocytosis \cite{liu2021lipid}, and SMCs in culture can rapidly efferocytose apoptotic SMCs \cite{bennett2016vascular}. This suggests that SMCs in plaques can acquire internalised lipid through ingestion of modLDL and apoptotic cells. In response to lipid loading, plaque SMCs may alter their phenotype to become macrophage-like cells \cite{allahverdian2012contribution}. In these SMC-derived macrophage-like cells (SDMs), SMC markers are suppressed and macrophage markers (including multiple pro-inflammatory genes) are activated \cite{shankman2015klf4}. In the absence of cellular lineage-tracing, it is therefore difficult to determine which cells that express macrophage markers are of SMC origin and which are of monocyte origin.

SDMs exhibit low expression of contractile markers and possess similar functions to MDMs, including innate immune signalling, phagocytosis, and efferocytosis \cite{bennett2016vascular,allahverdian2014contribution}. %Like MDMs, SDMs express a host of scavenger receptors and take on a foamy appearance as they ingest modLDL and accumulate other internalised lipids \cite{xiang2022smooth,bennett2016vascular,hashem2021stem}.
However, SDMs may be less effective than MDMs in clearing lipids and apoptotic cells from the lesion microenvironment, and they have a reduced phagocytic capacity compared to MDMs \cite{beyea2012oxysterol,wang2019smooth}.  They are also known to have a significantly reduced capacity to export internalised lipid to HDL meaning that, relative to MDMs, they retain a larger proportion of the lipid that they ingest \cite{allahverdian2012contribution}.
On the other hand, SDMs may proliferate much more rapidly than MDMs, to the extent that SMC-derived cells in plaques have been likened to tumour cells \cite{Pan_etal_2024}.

MDMs and SMC-derived cells can undergo programmed cell death in the plaque to become lipid-bearing apoptotic cells. If these apoptotic cells are not ingested and removed by living cells, they undergo secondary necrosis which leads to the formation of a necrotic core \cite{thorp2009mechanisms}. The necrotic core, a hallmark of advanced atherosclerosis, is associated with a high risk of thrombosis (blood clot formation) following plaque rupture \cite{back2019inflammation}.

%%%%%%%%%%%%%%%%%%%%%%%%%%%%%%%%%%%%%%%%%%%%%%%%%%%%%%%%%%%%%%%%%%%%%%%%%%%%%%%%%%%%%%
%%%The two paragraphs below have not been rewritten and probably need further attention.
%%%%%%%%%%%%%%%%%%%%%%%%%%%%%%%%%%%%%%%%%%%%%%%%%%%%%%%%%%%%%%%%%%%%%%%%%%%%%%%%%%%%%%%%
Mathematical modelling has increasingly been used to explore the cell and lipid dynamics of atherosclerotic plaque progression \cite{el2009mathematical,ougrinovskaia2010ode,cohen2014athero,bulelzai2014bifurcation,ford2019lipid,CHAMBERS2023108971,CHAMBERS2025112232}. Ford et al. \cite{ford2019lipid} developed a system of partial integro-differential equations to model the internalised lipid load distributions in live and apoptotic plaque macrophages. The authors explored how the trafficking of lipid between these populations contributes to the long-term formation of a necrotic core. Chambers et al.\ \cite{Chambers_etal_proliferation_2024} extended this model to include macrophage proliferation by assuming that parent cell internalised lipid is distributed between daughter cells during division. This provides a means to reduce internalised lipid loads in plaque cells.

Several existing models have explicitly focussed on the role of SMCs in plaques. Watson et al.\ \cite{watson2018two} developed a one-dimensional multiphase model to investigate fibrous cap formation by plaque SMCs. Their findings provide insight into how SMC behaviour can influence fibrous cap thickness, but the model does not consider the impact of phenotype switching from SMCs to SDMs. Pan et al.\ \cite{pan2021role} present a two-dimensional, hybrid discrete-continuous model that considers phenotype switching of plaque SMCs. The model identifies that the SMC-to-SDM transition can destabilise the plaque, but the phenotypic switching rate in this model is assumed to be constant and does not depend on SMC internalised lipid loads.  

In this paper, we propose an ordinary differential equation (ODE) model for plaque cell and lipid dynamics that incorporates phenotypic switching of SMCs to SDMs in response to SMC lipid loading. The model accounts for the population sizes and internalised lipid loads of MDMs, SMCs, and SDMs, and explores the implications of SMC phenotypic switching for long-term plaque fate.

The remaining sections of this paper are structured as follows. The mathematical model formulation is presented in Section \ref{s2}. Steady state analysis of a reduced model is presented in Section \ref{s3}, followed by numerical results of the full model in Section \ref{rsec2}. We conclude, in Section \ref{s4}, by discussing the findings of this study.
%%%%%%%%%%%%%%%%%%%%%%%%%%%%%%%%%%%%%%%%%%%%%%%%%%%%%%%%%%%%%%%%%%%%%%%
%%%%%%%%%%%%%%%%%%%%%%%%%%%%%%%%%%%%%%%%%%%%%%%%%%%%%%%%%%%

\section{Model formulation and definitions} \label{s2}
The model assumes that the plaque contains a dynamic mixture of LDL and HDL particles, MDMs, SMCs, SDMs, apoptotic cells, and necrotic core material. We let $\displaystyle{M(t)}$, $\displaystyle{C(t)}$, and $\displaystyle{S(t)}$ be time-dependent variables that represent the respective numbers of MDMs, SMCs, and SDMs. We define $\displaystyle{A_m(t)}$, $\displaystyle{A_c(t)}$, and $\displaystyle{A_s(t)}$ as the corresponding total lipid loads of these populations. The lipid inside each cell is assumed to include both the endogenous lipid $a_0$ (e.g., lipid in cell membranes), and lipid internalised from other sources. Additionally, $\displaystyle{A_p(t)}$ is the total lipid load of apoptotic cells, $\displaystyle{N(t)}$ is the lipid in the necrotic core, $\displaystyle{L(t)}$ is the total lipid on modLDL particles, and $\displaystyle{H(t)}$ is the total capacity of HDL particles to accept lipids offloaded by plaque cells. The equations that govern the evolution of these time-dependent quantities are presented below. %Here $\displaystyle{M,C,S,A_m,A_c,A_s,A_p,N,L}$ and $\displaystyle{H}$ are all non-negative real quantities.
\begin{figure}[h!]
	\centering
\boxed{\includegraphics[width=0.95\linewidth]{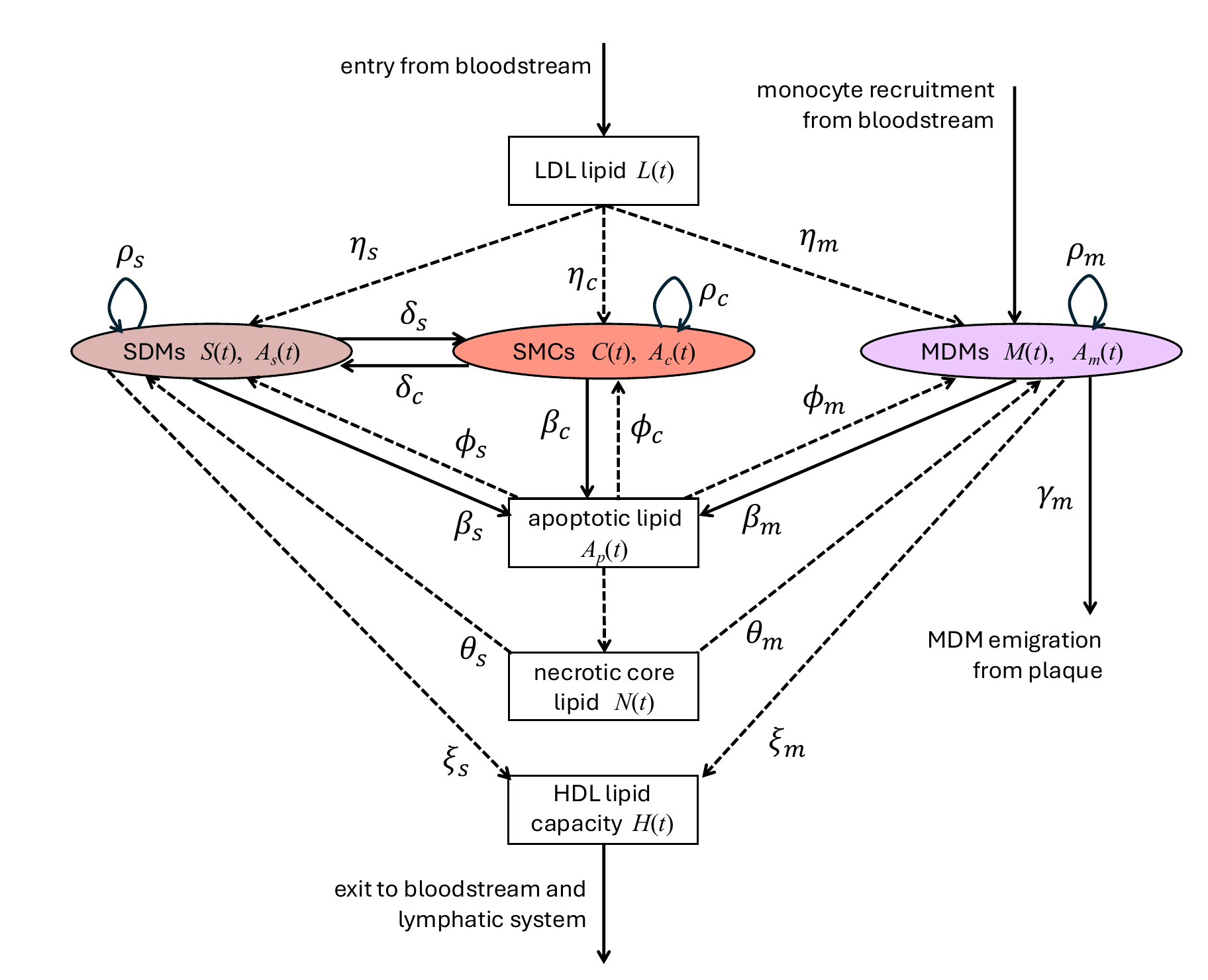}}
	\caption{Schematic diagram showing the interactions between cell and lipid species in the model plaque. Solid arrows indicate processes that affect the dynamics of both cells and lipids; dashed arrows indicate processes that affect only the dynamics of lipids. Labels on arrows correspond to the rate constants that appear in the differential equation model.}\label{SDp}
\end{figure}

\subsection{LDL and HDL}
We assume that lipid on native (unmodified) LDL particles enters the artery wall at a constant rate $\Lambda \sigma_L$, where $\Lambda$ denotes the rate of serum entry into the artery wall (volume per unit time), and $\sigma_L$ denotes the lipid mass on LDL particles per unit volume of serum. Once inside the artery wall, the LDL particles are rapidly modified to become modLDL. Lipid on modLDL particles is consumed by MDMs, SDMs, and SMCs at rates $\displaystyle{\eta_m, \eta_s}$, and $\displaystyle{\eta_c}$ (per cell per unit time), respectively. With these assumptions, the dynamics of $L(t)$ can be modelled by:
\begin{equation}
	\frac{dL}{dt} = \Lambda\sigma_L -(\eta_m M+\eta_sS+\eta_cC)L.\label{L1}
\end{equation}

We assume that HDL particles also enter the artery wall at constant rate $\Lambda$, and that these particles have a fixed capacity $\displaystyle{\sigma_H}$ for lipid acceptance (lipid mass capacity per unit volume of serum). Lipid is offloaded from MDMs and SDMs to HDL particles at the respective rates $\displaystyle{\xi_m}$ and $\displaystyle{\xi_s}$ (lipid mass per cell per HDL particle per unit time). The expression of genes to promote cholesterol exporter protein ATP-binding cassette transporter A1 (ABCA1), which are needed for lipid offload to HDL, is low in SDMs compared to MDMs, so that $\displaystyle{\xi_m>\xi_s}$ \cite{wang2019smooth,cai2019lncrna}. We assume that SMCs lack the machinery to offload lipid to HDL, and so $\xi_c \equiv 0$ \cite{allahverdian2012contribution}.
The lipid mass capacity of a HDL particle when it enters the artery wall is denoted $H_0$, and we assume that HDL particles become fully loaded with lipid before leaving the artery wall. With these assumptions, the model for HDL capacity is:
\begin{equation}
	\frac{dH}{dt} = \Lambda\sigma_H - (\xi_m M+\xi_sS)\frac{H}{H_0}.\label{H1}
\end{equation}

\subsection{Monocyte-derived macrophages (MDMs)}
We assume that monocytes enter the plaque from the bloodstream in response to pro-inflammatory cytokine signals, and then rapidly differentiate into macrophages. As cytokine signals are produced in response to modLDL accumulation \cite{hansson2005inflammation} and macrophage lipid loading \cite{allahverdian2014contribution,harrington2000role,reape1999chemokines}, we assume that the rate of recruitment of MDMs into the plaque is given by:
\begin{equation*}
	f(L,M,S,A_m,A_s) = \alpha_m\,\frac{L+\tau_m(A_m - a_0 M) +\tau_s(A_s -a_0 S)}{\kappa_m + L+\tau_m(A_m - a_0 M) +\tau_s(A_s -a_0 S)}.
\end{equation*}
Here, we assume that pro-inflammatory cytokine production in the plaque occurs proportionally to a total lipid stimulus $L+\tau_m(A_m - a_0 M) +\tau_s(A_s -a_0 S)$, and that the rate of monocyte recruitment is a saturating function of this stimulus with maximal recruitment rate $\alpha_m$. The magnitude of the lipid stimulus is a weighted sum of the total quantities of lipid in modLDL, and internalised in the MDM and SDM populations (with weightings $1$, $\tau_m$, and $\tau_s$, respectively). Half-maximal recruitment occurs when the value of the lipid stimulus is equal to $\kappa_m$.

In addition to recruitment, we assume that MDMs die via apoptosis, proliferate, and emigrate out of the plaque. The dynamics of $\displaystyle{M(t)}$ are therefore modelled by:
\begin{equation}
	\frac{dM}{dt} = \overbrace{f(L,M,S,A_m,A_s)}^{\text{recruitment}} + \overbrace{\rho_m M}^{\text{proliferation}} -\overbrace{\beta_m M}^{\text{apoptosis}}-\overbrace{\gamma_m M}^{\text{emigration}},\label{M}
\end{equation}
where $\displaystyle{\beta_m}$ is the MDM apoptosis rate, $\displaystyle{\gamma_m}$ is the MDM emigration rate, and $\displaystyle{\rho_m}$ is the MDM proliferation rate.
%\cite{Chambers_etal_proliferation_2024}.

%MDMs in the plaque are assumed to consume lipids by ingesting modLDL, necrotic core lipid, and lipid from apoptotic cells, including their lipid membranes and internalised lipid. MDMs may also offload lipids to HDL.
The dynamics of $\displaystyle{A_M(t)}$, the total lipid load of all MDMs, are modelled by:
\begin{multline}
\frac{dA_m}{dt} =  \overbrace{a_0f(L,M,S,A_m,A_s)}^{\text{recruitment}} +\overbrace{\eta_m LM}^{\text{modLDL ingestion}} - \overbrace{\frac{\xi_m HM}{H_0}}^{\text{offload to HDL}}  +\overbrace{\theta_m NM}^{\text{necrotic lipid consumption}} \\ +\underbrace{\phi_m A_pM}_{\text{efferocytosis}} + \underbrace{a_0\rho_m M}_{\text{proliferation}}- \underbrace{\beta_mA_m}_{\text{apoptosis}}-\underbrace{\gamma A_m}_{\text{emigration}}. \label{AM}
\end{multline}
The model assumes that MDMs can internalise lipid from modLDL, apoptotic cells, and necrotic cells at rates (per cell per time) $\eta_m$, $\phi_m$, and $\theta_m$, respectively. The MDM population acquires additional lipid when new cells enter the system. This corresponds to the endogenous lipid which is either carried into the plaque by newly-recruited cells, or \emph{de novo} synthesised during local proliferation \cite{Chambers_etal_proliferation_2024,Sawicki_etal_2019,Scaglia_etal_2014}. The total internalised lipid in MDMs is reduced when cells emigrate, die, or efflux lipid by offloading to HDL at rate $\frac{\xi_m}{H_0}$ (per cell per time).

\subsection{Smooth muscle cells (SMCs) and smooth muscle cell-derived macrophages (SDMs)}
A small number of SMCs, recruited into the plaque from the media, rapidly proliferate to colonise the region beneath the endothelium \cite{misra2018integrin}. Exposure to lipids can stimulate these SMCs to differentiate into macrophage-like cells \cite{allahverdian2014contribution}, but this phenotypic change can be reversed by lipid offloading \cite{vengrenyuk2015cholesterol}. Therefore, we model the dynamics of $C(t)$ by the equation:
\begin{multline}
	\frac{dC}{dt} =
	\overbrace{\rho_c \!\left(1-\frac{C}{C_{\max}}\right)C}^{\text{proliferation}}
	- \overbrace{\beta_c C}^{\text{apoptosis}}
	- \overbrace{\delta_c\,\frac{\left(\tfrac{A_c}{C}-a_0\right)^n C}{\alpha_c^n+\left(\tfrac{A_c}{C}-a_0\right)^n}}^{\text{SMC-to-SDM switch}} \\ 
	+ \overbrace{\delta_s\!\left(1-\frac{\left(\tfrac{A_s}{S}-a_0\right)^n}{\alpha_s^n+\left(\tfrac{A_s}{S}-a_0\right)^n}\right)S}^{\text{SDM-to-SMC switch}}.
	\label{C}
\end{multline}
The first term represents the proliferation of SMCs in the cap region, where $\displaystyle{\rho_c}$ denotes the maximum proliferation rate and $\displaystyle{C_{max}}$ denotes the carrying capacity. Physically, the value of the carrying capacity is assumed to reflect the availability of growth factors and space proximal to the endothelium \cite{hedin2004control,mehrhof2005regulation}. Mathematically, a growth limiting term is required to prevent unbounded SMC growth in the absence of phenotype switching. The second term in equation \eqref{C} represents SMC apoptosis at rate $\displaystyle{\beta_c}$.

We assume that the likelihood of an SMC becoming a SDM increases with lipid loading, and that the likelihood of a SDM reverting to a SMC decreases with lipid loading. This is modelled by the final two terms in \eqref{C}, where $\displaystyle{\delta_c}$ is the maximum SMC-to-SDM switching rate, and $\displaystyle{\delta_s}$ is the maximum SDM-to-SMC switching rate. The overall switching rates are assumed to be functions of the ingested lipid per cell. That is, $(A_c/C - a_0)$ and $(A_s/S - a_0)$ for the forward and backward switching, respectively. We use a Hill function formulation to express the fact that the SMC-to-SDM phenotype switch is unlikely if average SMC lipid is low, and the SDM-to-SMC phenotype switch is unlikely if average SDM lipid is high. The exponent $n\geqslant1$ controls the sharpness of the switch. The average ingested lipid loads for a half-maximal switching rate are $\alpha_c$ and $\alpha_s$, where we assume $\alpha_c>\alpha_s$.

Plaque SMCs express various receptors that mediate modLDL uptake \cite{liu2021lipid}. SMCs in culture also rapidly ingest apoptotic SMCs \cite{bennett2016vascular}. We therefore assume that the total lipid load of the SMC population $\displaystyle{A_c(t)}$ has dynamics given by:

\begin{multline}
	\frac{dA_c}{dt} =
	\overbrace{\eta_c L C}^{\text{modLDL ingestion}}
	+ \overbrace{\phi_c A_p C}^{\text{efferocytosis}}
	+ \overbrace{a_0 \rho_c \!\left(1-\frac{C}{C_{\max}}\right)C}^{\text{proliferation}}
	- \overbrace{\beta_c A_c}^{\text{apoptosis}} \\
	- \underbrace{\delta_c\,\frac{\left(\tfrac{A_c}{C}-a_0\right)^n A_c}{\alpha_c^n+\left(\tfrac{A_c}{C}-a_0\right)^n}}_{\text{SMC-to-SDM switch}}
	+ \underbrace{\delta_s\!\left(1-\frac{\left(\tfrac{A_s}{S}-a_0\right)^n}{\alpha_s^n+\left(\tfrac{A_s}{S}-a_0\right)^n}\right)A_s}_{\text{SDM-to-SMC switch}},\label{AC}
\end{multline}
where $\eta_c$ and $\phi_c$ are the rates (per cell per time) of lipid ingestion from modLDL and apoptotic cells, respectively. As for MDMs, we assume that proliferating SMCs \emph{de novo} generate endogenous lipid for their daughter cells. This produces the proliferation term in equation \eqref{AC}. The remaining terms in equation \eqref{AC} represent the lipid lost to the apoptotic lipid pool upon SMC apoptosis, the lipid transferred to the SDM population upon SMC-to-SDM phenotypic switching, and the lipid regained from the SDM population upon SDM-to-SMC phenotypic switching. 

The SDM population dynamics share features of the MDM and SMC populations. The SDMs are subject to phenotypic switching, proliferation (at rate $\rho_s$), and apoptosis (at rate $\beta_s$), and we assume that they do not emigrate \cite{liu2021lipid,bennett2016vascular}. The SDM dynamics $S(t)$ are modelled by the equation:
\begin{equation}
	\frac{dS}{dt} =
\overbrace{\delta_c\,\frac{\left(\tfrac{A_c}{C}-a_0\right)^n C}{\alpha_c^n+\left(\tfrac{A_c}{C}-a_0\right)^n}}^{\text{SMC-to-SDM switch}}
	+ \overbrace{\rho_s S}_{\text{proliferation}}
	- \overbrace{\beta_s S}^{\text{apoptosis}}
	- \overbrace{\delta_s\!\left(1-\frac{\left(\tfrac{A_s}{S}-a_0\right)^n}{\alpha_s^n+\left(\tfrac{A_s}{S}-a_0\right)^n}\right)S}^{\text{SDM-to-SMC switch}}.
	\label{S}
\end{equation}

We assume that SDMs ingest lipid from modLDL, apoptotic cells, and necrotic material at rates (per cell per time) $\eta_s$, $\phi_s$, and $\theta_s$, respectively. We also assume that SDMs can offload ingested lipid to HDL at rate $\frac{\xi_s}{H_0}$ (per cell per time). The dynamics of the total lipid load of the SDM population $A_s(t)$ therefore follow:

\begin{multline}
    \frac{dA_s}{dt} =
	\overbrace{\eta_s L S}^{\text{modLDL ingestion}}
	- \overbrace{\tfrac{\xi_s H S}{H_0}}^{\text{HDL offloading}}
	+ \overbrace{\theta_s N S}^{\text{necrotic lipid consumption}}
	+ \overbrace{\phi_s A_p S}^{\text{efferocytosis}} \\
	+ \underbrace{\delta_c\,\frac{\left(\tfrac{A_c}{C}-a_0\right)^n A_c}{\alpha_c^n+\left(\tfrac{A_c}{C}-a_0\right)^n}}_{\text{SMC-to-SDM switch}}
	+ \underbrace{a_0 \rho_s S}_{\text{proliferation}}
	- \underbrace{\beta_s A_s}_{\text{apoptosis}}
	- \underbrace{\delta_s\!\left(1-\frac{\left(\tfrac{A_s}{S}-a_0\right)^n}{\alpha_s^n+\left(\tfrac{A_s}{S}-a_0\right)^n}\right)A_s}_{\text{SDM-to-SMC switch}}.\label{AS}
\end{multline}
In the above equation, we assume that the parameters $\eta_s$, $\phi_s$ and $\theta_s$ all have smaller values than the corresponding parameters ($\eta_m$, $\phi_m$, $\theta_m$) for MDMs.

\subsection{Apoptotic lipid and necrotic core lipid}
We assume that all cell types contribute to a single class of apoptotic cells when they die. As such, a cell does not “know” whether the apoptotic lipid it ingests was originally contained in a MDM, SMC, or SDM. All apoptotic cells undergo secondary necrosis at rate $\displaystyle{\nu}$ if not ingested by another live cell. The total mass of lipid in the apoptotic lipid pool $A_p(t)$ is therefore governed by the equation:
\begin{equation}
	\frac{dA_p}{dt} = \beta_m A_m + \beta_s A_s + \beta_c A_c - \left(\phi_m M + \phi_s S + \phi_c C\right)A_p - \nu A_p,\label{P}
\end{equation}
where the first three terms model the accumulation of apoptotic lipid due to plaque cell apoptosis, the following three terms model plaque cell ingestion of apoptotic lipid, and the final term models the loss of apoptotic lipid due to secondary necrosis of apoptotic cells.

The corresponding dynamics of the necrotic lipid mass $N(t)$ are given by the equation:
\begin{equation}
	\frac{dN}{dt} = \nu A_p - \left(\theta_m M + \theta_s S\right)N,\label{N}
\end{equation}
where the first term represents necrotic lipid generation due to secondary necrosis of apoptotic cells, and the final two terms represent necrotic lipid consumption by MDMs and SDMs. It is assumed that SMCs do not consume necrotic lipid as plaque cell necrosis occurs mostly in the core of the plaque, which is distal to the cap region where SMCs accumulate.

\subsection{Initial conditions and model parameterisation}
At time $t=0$, the equations \eqref{L1}--\eqref{N} are subject to the initial conditions:
\begin{equation}            L(0)=H(0)=M(0)=A_m(0)=C(0)=A_c(0)=S(0)=A_s(0)=A_p(0)=N(0)=0.
\end{equation}
From these zero initial conditions, plaque formation is initiated by an influx of LDL lipid which, in turn, stimulates the recruitment of MDMs. No SMCs, nor SDMs, enter the plaque at this stage. Consistent with experimental observations, we assume that SMCs first enter the plaque several weeks after the MDMs \cite{misra2018integrin}. Thus, at time $t=t_{c}>0$, we introduce a small population of SMCs containing only their endogeneous lipid by setting:
\begin{equation}
C(t_c)=C_{init}, \,\text{ and }\, A_c(t_c)=a_0\,C_{init}.
\end{equation}
The subsequent growth of this SMC population by proliferation provides a potential source of SDMs via phenotypic switching. 

See Table \ref{tab1} for a summary of the dimensional parameters used in the model.
\begin{sidewaystable}
			\small
			\centering
			\begin{tabular}{clccc}
				\hline
				Parameter& Description&Value & Unit& Source\\
				\hline
				$\alpha_m$ & MDMs maximum recruitment rate & $5.4$ & cell/hour & \cite{swirski2006monocyte}\\
				$\kappa_m$ & Lipid stimulus for half-maximal MDM recruitment & $5\times 10^{-8}$ & g &\\
                $\rho_m$ & MDMs proliferation rate & $0.0005$ & per hour & \cite{robbins2013local,tang2015inhibiting}\\   
                $\beta_m$ & MDMs apoptosis rate & $0.002$ & per hour & \cite{yona2013fate}\\
                $\gamma$ & MDMs emigration rate & $0.0015$ & per hour & \cite{williams2018limited,lee2019sirt1}\\
				$a_0$ & Mass of endogenous lipid in each cell & $26.6\times 10^{-12}$ & g/cell & \cite{sokol1991changes,cooper2022cell}\\
				$\rho_c$ & SMCs proliferation rate & $0.016$ & per hour & \cite{misra2018integrin}%jenkins2011local}
                \\ 
				$C_{max}$ & SMCs maximum carrying capacity & $750$ & cells &\\
				$\beta_c$ & SMCs apoptosis rate & $0.0004$ & per hour & %\cite{bennett1995apoptosis}
                \\ 
				$\delta_c$ & SMCs maximum switching rate to SDMs & $0.006$ & per hour & %\cite{shankman2015klf4}
                \\
				$\alpha_c$ & SMC average ingested lipid load for half-maximal switching to SDM & $2a_0$ & g/cell &\\
                $\rho_s$ & SDMs proliferation rate & $0.625\rho_c$ & per hour & \cite{misra2018integrin}\\
                $\beta_s$ & SDMs apoptosis rate & $1.06\rho_s$ & per hour & \cite{shankman2015klf4}\\
				$\delta_s$ & SDMs maximum switching rate to SMCs & $\delta_c$ & per hour &\\
                $\alpha_s$ & SDM average ingested lipid load for half-maximal switching to SMC & $0.5a_0$ & g/cell &\\  
				$\Lambda$ & Rate of serum entry into artery wall & $5\times10^{-4}$ & $\mu$L/hour & \cite{nielsen1996transfer}\\
				$\sigma_L$ & Lipid content of LDL particles per unit volume of serum & $8\times10^{-7}$ & g/$\mu$L & \cite{lee2012characteristics,orlova1999three}\\
				$\sigma_H$ & Lipid capacity of HDL particles per unit volume of serum & $5\times10^{-7}$ & g/$\mu$L & \cite{casula2021hdl} \\
				$H_0$ & Maximum lipid capacity of single HDL particle & $5\times10^{-17}$ & g/HDL particle & \cite{kontush2015structure,matyus2015hdl} \\
				$\eta_m$ & modLDL consumption rate by MDMs  & $1\times10^{-6}$ & per cell per hour &\\
                $\eta_c$ & modLDL consumption rate by SMCs & $0.4\eta_m$ & per cell per hour &\\
				$\eta_s$ & modLDL consumption rate by SDMs  & $0.75\eta_m$ & per cell per hour &\cite{vengrenyuk2015cholesterol}\\
				$\xi_m$ & MDM offloading rate of lipid to HDL & $1\times10^{-23}$ & g/cell per HDL particle per hour &\\
				$\xi_s$ & SDM offloading rate of lipid to HDL & $0.25\xi_m$ & g/cell per HDL particle per hour &\cite{vengrenyuk2015cholesterol}\\
				$\phi_m$ & Apoptotic lipid consumption rate by MDMs & $1\times10^{-5}$ & per cell per hour &\cite{ford2019lipid,ford2019efferocytosis}\\
                $\phi_c$ & Apoptotic lipid consumption rate by SMCs & $0.2\phi_m$ & per cell per hour &\\
				$\phi_s$ & Apoptotic lipid consumption rate by SDMs & $0.25\phi_m$ & per cell per hour &\cite{vengrenyuk2015cholesterol}\\
				$\theta_m$ & Necrotic lipid consumption rate by MDMs & $3.6\times10^{-6}$ & per cell per hour & \cite{schrijvers2005phagocytosis}\\
				$\theta_s$ & Necrotic lipid consumption rate by SDMs & $0.25\theta_m$ & per cell per hour & \cite{vengrenyuk2015cholesterol}\\
				$\nu$ & Secondary necrosis rate & $0.05$ & per hour & \cite{ford2019lipid,Saraste_et_al_2000}\\
                $t_c$ & Time of SMC entry into plaque & $850$ & hours & \cite{misra2018integrin}\\
                $C_{init}$ & SMC population size at time $t=t_c$ & $1.35$ & cells & \cite{misra2018integrin}\\
				\hline
			\end{tabular}
			\caption{Summary of dimensional model parameters.}
			\label{tab1}
\end{sidewaystable}

\subsection{Nondimensionalisation}
We rewrite the model equations \eqref{L1}--\eqref{N} in terms of the following dimensionless variables, denoted with tildes:
\begin{align*}
	\Tilde{t}:= \beta_m t,\quad \Tilde{C}:=\frac{\beta_m}{\alpha_m}C,\quad \Tilde{S}:=\frac{\beta_m}{\alpha_m}S,\quad\Tilde{M}:=\frac{\beta_m}{\alpha_m}M,\quad \Tilde{L}:=\frac{\beta_m}{a_0\alpha_m}L,\quad \Tilde{H}:=\frac{\beta_m}{a_0\alpha_m}H, \nonumber\\ \Tilde{A_c}:=\frac{\beta_m}{a_0\alpha_m}A_c, \quad\Tilde{A_s}:=\frac{\beta_m}{a_0\alpha_m}A_s\quad \Tilde{A}_m:=\frac{\beta_m}{a_0\alpha_m}A_m, \quad \Tilde{A_p}:=\frac{\beta_m}{a_0\alpha_m}A_p, \quad 
	\Tilde{N}:=\frac{\beta_m}{a_0\alpha_m}N.
\end{align*}
The corresponding dimensionless parameters are defined in Table \ref{tab3}. 

Dropping the tildes for notational convenience, we have the following non-dimensional ODE system:
\begin{subequations}
	\footnotesize
	\begin{align}
	\frac{dL}{dt} &=\sigma_L -(\eta_mM+\eta_sS+\eta_cC)L\label{Leqn}\\
		\frac{dH}{dt} &=\sigma_H -(\zeta_mM+\zeta_sS)H\label{Heqn}\\
        \frac{dC}{dt} &= \rho_c\left(1-\frac{C}{C_0}\right)C - \beta_c C - \delta_c\,\frac{\left(\tfrac{A_c}{C}-1\right)^n C}{\alpha_c^n+\left(\tfrac{A_c}{C}-1\right)^n} + \delta_s\left(1-\frac{\left(\tfrac{A_s}{S}-1\right)^n}{\alpha_s^n+\left(\tfrac{A_s}{S}-1\right)^n}\right)S\label{Ceqn} \\
		\frac{dA_c}{dt} & = \eta_c LC+\Phi_c A_pC+ \rho_c\left(1-\frac{C}{C_0}\right)C - \beta_c A_c -\delta_c\,\frac{\left(\tfrac{A_c}{C}-1\right)^n A_c}{\alpha_c^n+\left(\tfrac{A_c}{C}-1\right)^n} + \delta_s\left(1-\frac{\left(\tfrac{A_s}{S}-1\right)^n}{\alpha_s^n+\left(\tfrac{A_s}{S}-1\right)^n}\right)A_s \label{Aceqn}\\
        \frac{dS}{dt} & = \delta_c\,\frac{\left(\tfrac{A_c}{C}-1\right)^n C}{\alpha_c^n+\left(\tfrac{A_c}{C}-1\right)^n} + \rho_s S - \beta_s S -  \delta_s\left(1-\frac{\left(\tfrac{A_s}{S}-1\right)^n}{\alpha_s^n+\left(\tfrac{A_s}{S}-1\right)^n}\right)S\label{Seqn}\\
        \frac{dA_s}{dt}& = \left(\eta_sL-\zeta_sH+\Phi_s A_p +\Theta_s N + \rho_s \right)S - \beta_s A_s + \delta_c\,\frac{\left(\tfrac{A_c}{C}-1\right)^n A_c}{\alpha_c^n+\left(\tfrac{A_c}{C}-1\right)^n} - \delta_s\left(1-\frac{\left(\tfrac{A_s}{S}-1\right)^n}{\alpha_s^n+\left(\tfrac{A_s}{S}-1\right)^n}\right)A_s\label{Aseqn}\\
		\frac{dM}{dt} &= \frac{L+\tau_m(A_m-M)+\tau_s(A_s-S)}{\Gamma+L+\tau_m(A_m-M)+\tau_s(A_s-S)} + \rho_m M - M - 	\gamma M\label{Meqn}\\
		\frac{dA_m}{dt} &= \frac{L+\tau_m(A_m-M)+\tau_s(A_s-S)}{\Gamma+L+\tau_m(A_m-M)+\tau_s(A_s-S)} +\left( \eta_mL-\zeta_mH+\Phi_m A_p +\Theta_m N + \rho_m \right)M - \left(1+ \gamma\right)A_m\label{Ameqn}\\
		\frac{dA_p}{dt} &= A_m + \beta_s A_s + \beta_c A_c - \left(\Phi_m M + \Phi_s S + \Phi_c C\right)A_p - \nu A_p\label{Apeqn}\\
		\frac{dN}{dt} &= \nu A_p - \left(\Theta_m M + \Theta_s S\right)N\label{Neqn}
	\end{align} \label{MS}
\end{subequations}

These equations are once again subject to zero initial conditions at time $t=0$. SMCs are introduced into the system at time $t=t_c>0$ according to the following conditions:
\begin{equation}
C(t_c)=A_c(t_c)=C_{init}.
\end{equation}
\begin{table}	
	%\footnotesize
	\centering
	\scriptsize
	\renewcommand{\arraystretch}{2.0}
	\begin{tabularx}{\textwidth}{cclc}
		\FL
		Parameter&Definition& Description&Approx. Baseline Value \ML[0.08em]
		$\displaystyle{\tilde{\rho}_c}$ &$\displaystyle{\frac{\rho_c}{\beta_m}}$ &SMCs proliferation rate&8 \NN
		$\displaystyle{\tilde{C_0}}$ &$\displaystyle{\frac{\beta_mC_{max}}{\alpha_m}}$ &SMCs maximum carrying capacity&0.28 \NN
		$\displaystyle{\tilde{\beta}_c}$ & $ \displaystyle{\frac{\beta_c}{\beta_m}}$&SMCs apoptosis rate&0.2 \NN
		$\displaystyle{\tilde{\delta}_c}$ & $\displaystyle{\frac{\delta_c}{\beta_m}}$&Maximum SMC-to-SDM phenotype switching rate &3 \NN
		$\displaystyle{\tilde{\alpha}_c}$ &$\displaystyle{\frac{\alpha_c}{a_0}}$&Ingested lipid load for half-maximal SMC-to-SDM switching&2 \NN
		$\displaystyle{\tilde{\delta}_s}$ & $\displaystyle{\frac{\delta_s}{\beta_m}}$ &Maximum SDM-to-SMC phenotype switching rate&3 \NN
		$\displaystyle{\tilde{\alpha}_s}$ & $\displaystyle{\frac{\alpha_s}{a_0}}$&Ingested lipid load for half-maximal SDM-to-SMC switching&0.5 \NN
		$\displaystyle{\tilde{\sigma}_L}$ &  $\displaystyle{\frac{\Lambda \sigma_L}{a_0\alpha_m}}$&Net influx rate of lipids on LDL&2.78 \NN
		$\displaystyle{\tilde{\sigma}_H}$ &  $\displaystyle{\frac{\Lambda \sigma_H }{a_0\alpha_m}}$&Net influx rate of HDL lipid efflux capacity&1.74 \NN
		$\displaystyle{\Phi_m}$ & $\displaystyle{\frac{\phi_m\alpha_m}{\beta_m^2}}$& MDM apoptotic lipid consumption rate &13.5 \NN
		$\displaystyle{\Phi_s}$ & $\displaystyle{\frac{\phi_s\alpha_m}{\beta_m^2}}$& SDM apoptotic lipid consumption rate &3.38 \NN
		$\displaystyle{\Phi_c}$ & $\displaystyle{\frac{\phi_c\alpha_m}{\beta_m^2}}$& SMC apoptotic lipid consumption rate &2.7 \NN
		$\displaystyle{\Theta_m}$ &$\displaystyle{\frac{\theta_m\alpha_m}{\beta_m^2}}$& MDM necrotic lipid consumption rate&4.86 \NN
		$\displaystyle{\Theta_s}$ &$\displaystyle{\frac{\theta_s\alpha_m}{\beta_m^2}}$& SDM necrotic lipid consumption rate&1.22 \NN
		$\displaystyle{\tilde{\eta}_m}$&$\displaystyle{\frac{\alpha_m\eta_m}{\beta_m^2}}$ & MDM modLDL consumption rate&$1.35$ \NN
		$\displaystyle{\tilde{\eta}_s}$&$\displaystyle{\frac{\alpha_m\eta_s}{\beta_m^2}}$ & SMC modLDL consumption rate&$1.01$ \NN
		$\displaystyle{\tilde{\eta}_c}$&$\displaystyle{\frac{\alpha_m\eta_c}{\beta_m^2}}$ & SDM modLDL consumption rate&$0.54$ \NN
		$\displaystyle{\zeta_m}$&$\displaystyle{\frac{\alpha_m\xi_m}{H_0\beta_m^2}}$ & MDM offloading rate to HDL&$0.27$ \NN
		$\displaystyle{\zeta_s}$&$\displaystyle{\frac{\alpha_m\xi_s}{H_0\beta_m^2}}$ & SDM offloading rate to HDL&$0.07$ \NN
		$\displaystyle{\tilde{\nu}}$&$\displaystyle{\frac{\nu}{\beta_m}}$ & Secondary necrosis rate&25 \NN
		$\displaystyle{\tilde{\rho}_s}$ & $\displaystyle{\frac{\rho_s}{\beta_m}}$&SDM proliferation rate&5 \NN
		$\displaystyle{\tilde{\beta}_s}$ & $ \displaystyle{\frac{\beta_s}{\beta_m}}$&SDM apoptosis rate&5.3 \NN
		$\displaystyle{\tilde{\Gamma}}$ & $\displaystyle{\frac{\kappa_m\beta_m}{a_0\alpha_m}}$&Lipid stimulus for half-maximal MDM recruitment&0.70 \NN
		$\displaystyle{\tilde{\rho}_m}$ & $\displaystyle{\frac{\rho_m}{\beta_m}}$&MDM proliferation rate&0.25 \NN
		$\displaystyle{\tilde{\gamma}}$ &$\displaystyle{\frac{\gamma}{\beta_m}}$ &MDM emigration rate&0.75  \NN
        $\displaystyle{\tilde{t}_c}$ &$\displaystyle{\beta_m t_c}$ &Time of SMC entry into plaque&1.7  \NN
        $\displaystyle{\tilde{C}_{init}}$ &$\displaystyle{\frac{\beta_mC_{init}}{\alpha_m}}$ &SMC population at time $\tilde{t}=\tilde{t}_c$&$5\times10^{-4}$ \NN
		$\tau_m$ & -& Weighting of MDM ingested lipid stimulus for MDM recruitment&1\NN
		$\tau_s$ &- &Weighting of SDM ingested lipid stimulus for MDM recruitment&1\NN
        $n$ &-&Hill coefficient in switching functions&4\LL
	\end{tabularx}
	\caption{Summary of dimensionless model parameters}
	\label{tab3}
\end{table}

\section{Reduction to two equations---SMC dynamics} \label{s3}
In this section, we set $\delta_s = 0$ and assume that SMC lipid uptake occurs at a constant rate $\Pi_c>0$, which is independent of $L$ and $A_p$. This decouples equations \eqref{Ceqn} and \eqref{Aceqn} from the rest of the system \eqref{MS} and allows us to analyse, in detail, the long-term dynamics of the cap SMCs, $\displaystyle{C}$, and their total lipid load, $\displaystyle{A_c}$. The non-dimensional equations for $C(t)$ and $A_c(t)$ in this case are:
\begin{subequations}
    \begin{align}
        \frac{dC}{dt} &= \rho_c C\left(1-\frac{C}{C_0}\right) - \beta_c C -  \delta_c C \frac{ \left(\frac{A_c}{C}-1\right)^n}{\alpha_c^n+\left(\frac{A_c}{C}-1\right)^n},\label{dCdt_init} \\
	    \frac{dA_c}{dt} &= \Pi_cC + \rho_c C\left(1-\frac{C}{C_0}\right) - \beta_c A_c - \delta_c A_c\frac{ \left(\frac{A_c}{C}-1\right)^n}{\alpha_c^n+\left(\frac{A_c}{C}-1\right)^n}.\label{dACdt}
    \end{align}\label{system1}
\end{subequations}
The steady state analysis of the above system can be simplified by replacing the equation for $A_c$ with the corresponding equation for the average SMC lipid load $\bar{A}_c = \frac{A_c}{C}\geq1$. In what follows, we therefore study the system:
\begin{subequations}
    \begin{align}
        \frac{dC}{dt} &= \rho_cC\left(1-\frac{C}{C_0}\right) - \beta_c C -  \delta_cC\frac{ \left(\bar{A}_c-1\right)^n}{\alpha_c^n+\left(\bar{A}_c-1\right)^n},\label{dCdt} \\
	    \frac{d\bar{A}_c}{dt} &= \Pi_c - \rho_c\left(1-\frac{C}{C_0}\right)\left(\bar{A}_c-1\right).\label{dAbardt}
    \end{align}\label{system2}
\end{subequations}
The system \eqref{system2} has two steady state solutions. The first is:
\begin{equation}
    C^* = 0, \;\text{ and }\; \bar{A}_c^* = 1+\frac{\Pi_c}{\rho_c}.   \label{SS_triv}
\end{equation}
This corresponds to the trivial steady state $C^*=0$, $A_c^*=0$ of the system \eqref{system1}, and indicates that, on approach to this steady state, the average ingested lipid load maintains a non-zero value. Assuming that $C^*\neq0$, the second steady state solution of \eqref{system2} is:
\begin{equation}
	C^* = C_0 \left(1-\frac{\Pi_c}{k\rho_c}\right), \;\text{ and }\; \bar{A}_c^* = 1+k,   \label{SS_non_triv}
\end{equation}
where $k>0$ is the real, positive solution of the equation: 
\begin{equation}
	\left(\beta_c+\delta_c\right)k^{n+1}-\Pi_ck^n+\alpha_c^n\beta_ck-\alpha_c^n\Pi_c = 0.   \label{k_eqn}
\end{equation}
In the special case $n=1$, equation \eqref{k_eqn} is a quadratic in $k$ that has solution:
\begin{equation}
	k = \frac{\Pi_c-\beta_c\alpha_c+\sqrt{\left(\Pi_c+\beta_c\alpha_c\right)^2+4\delta_c\Pi_c\alpha_c}}{2\left(\beta_c+\delta_c\right)}. \label{k_n1}
\end{equation}
The steady state solution \eqref{SS_non_triv} is positive and, hence, physical provided that:
\begin{equation}
	\rho_c>\frac{\Pi_c}{k},\label{cond1}
\end{equation}
where:
\begin{equation}
	\frac{\Pi_c}{k}=\beta_c+\delta_c\frac{k^n}{\alpha_c^n+k^n},\label{cond2}
\end{equation}
such that $\displaystyle{\frac{\Pi_c}{k}}\in\left(\beta_c\,,\beta_c+\delta_c\right)$.

To determine the stability of the steady state solutions \eqref{SS_triv} and \eqref{SS_non_triv}, we derive the Jacobian matrix $J\left(C,\bar{A}_c\right)$ of the system \eqref{system2}. The Jacobian matrix is:
\begin{equation}
    J\left(C,\bar{A}_c\right) = \begin{bmatrix}
    \rho_c\left(1-\frac{2C}{C_0}\right) - \beta_c -\delta_c\frac{ \left(\bar{A}_c-1\right)^n}{\alpha_c^n+\left(\bar{A}_c-1\right)^n} & 
    -\frac{n\alpha_c^n \delta_cC\left(\bar{A}_c-1\right)^{n-1}}{\left(\alpha_c^n+\left(\bar{A}_c-1\right)^n\right)^2} \\[1.5em]
    \frac{\rho_c}{C_0}\left(\bar{A}_c-1\right) &
    \rho_c\left(\frac{C}{C_0}-1\right)
    \end{bmatrix}.
\end{equation}
Evaluating the Jacobian at the steady state \eqref{SS_non_triv} gives:
\begin{equation}
    J\left(C^*,\bar{A}_c^*\right) = \begin{bmatrix}
    \frac{\Pi_c}{k}-\rho_c & 
    -\frac{n\alpha_c^n \delta_ck^{n-1}C_0\left(1-\frac{\Pi_c}{k\rho_c}\right)}{\left(\alpha_c^n+k^n\right)^2} \\[1.5em]
    \frac{k\rho_c}{C_0} &
    -\frac{\Pi_c}{k}
    \end{bmatrix}.
\end{equation}
Here, we have $\text{tr}J\left(C^*,\bar{A}_c^*\right)<0$, and $\text{det}J\left(C^*,\bar{A}_c^*\right)>0$, provided that $\rho_c>\frac{\Pi_c}{k}$. Hence, this steady state solution is stable when it is physical because both eigenvalues of the Jacobian have negative real parts. Evaluating the Jacobian at the steady state \eqref{SS_triv} gives:
\begin{equation}
    J\left(C^*,\bar{A}_c^*\right) = \begin{bmatrix}
    \rho_c - \beta_c -\delta_c\frac{\left(\frac{\Pi_c}{\rho_c}\right)^n}{\alpha_c^n+\left(\frac{\Pi_c}{\rho_c}\right)^n} & 
    0 \\[1.5em]
    \frac{\Pi_c}{C_0} &
    -\rho_c
    \end{bmatrix}.
\end{equation}
This matrix has the real eigenvalues $\lambda_1=\rho_c - \beta_c -\delta_c\frac{\left(\frac{\Pi_c}{\rho_c}\right)^n}{\alpha_c^n+\left(\frac{\Pi_c}{\rho_c}\right)^n}$ and $\lambda_2=-\rho_c$. The steady state solution is therefore an unstable saddle when $\lambda_1>0$, and a stable node when $\lambda_1<0$. Using the condition \eqref{cond1} and equation \eqref{cond2}, we deduce that $\lambda_1>0$ when the steady state \eqref{SS_non_triv} is physical, and $\lambda_1<0$ otherwise. This analysis implies that, in an environment where SMC proliferation is not sufficiently fast to dominate the combined effects of apoptosis and phenotype switching, the SMC population that is required to form the cap will become extinct in the long-term.

\begin{figure}%[h!]
	\centering
    \includegraphics[height=6cm]{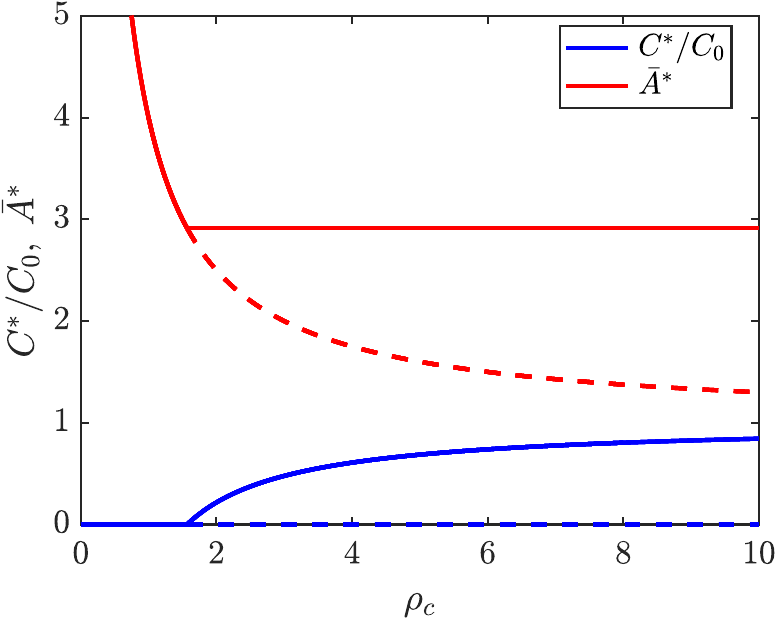}
	\caption{Bifurcation diagram showing the physical steady state solutions \eqref{SS_triv} and \eqref{SS_non_triv} as functions of $\rho_c$. Blue lines show $C^*$ (normalised by the carrying capacity $C_0$), and red lines show $\bar{A}^*$. Stable branches are indicated by solid lines, and unstable branches by dashed lines. We set $\Pi_c = 3$, and the values of all other relevant parameters are given in Table \ref{tab3}.
    }\label{bif_rhoc}
\end{figure}

Figure \ref{bif_rhoc} plots the stable and unstable branches of the steady state solutions \eqref{SS_triv} and \eqref{SS_non_triv} as functions of the SMC proliferation rate $\rho_c$ for $\Pi_c=3$ and all other parameters at their baseline values. The plot shows that long-term extinction of the SMC population requires $\rho_c<1.57$, while, at the baseline proliferation rate ($\rho_c=8$), the steady state SMC population is relatively close to its theoretical maximum.

\begin{figure}%[h!]
	\centering
	\begin{subfigure}[b]{0.49\textwidth}
		\centering	
		\includegraphics[height=5cm]{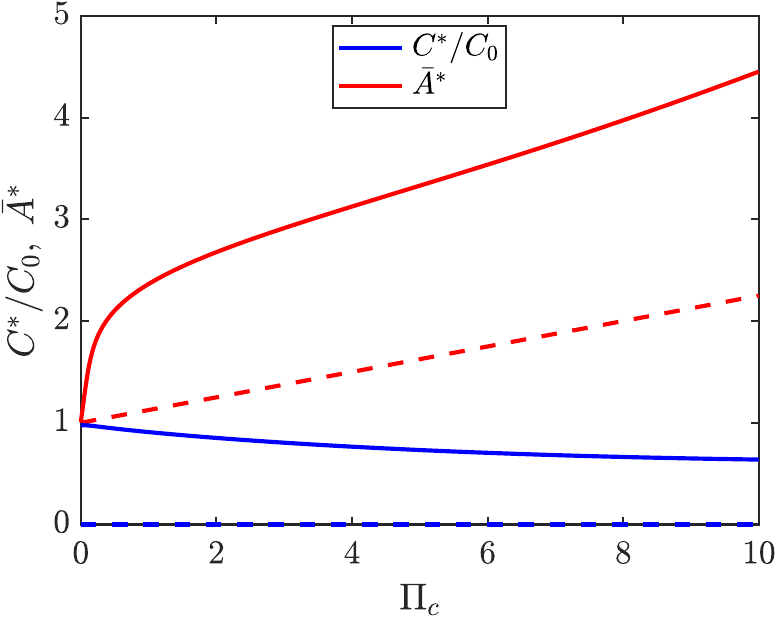}
		\caption{} \label{bif_Pic}
	\end{subfigure}
	\begin{subfigure}[b]{0.49\textwidth}
		\centering	
		\includegraphics[height=5cm]{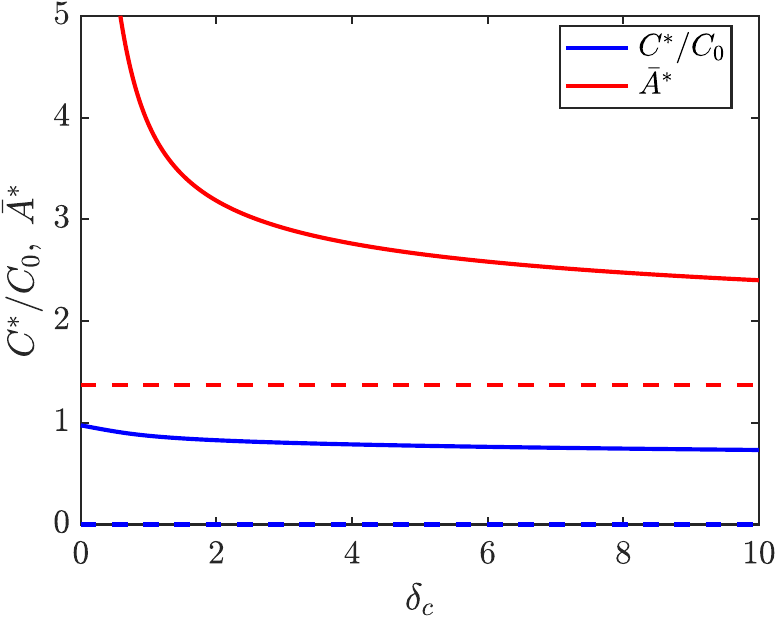}
		\caption{} \label{bif_deltac}
	\end{subfigure}
	\caption{Bifurcation diagrams showing the steady state solutions \eqref{SS_triv} and \eqref{SS_non_triv} as functions of (a) $\Pi_c$ and (b) $\delta_c$. Blue lines show $C^*$ (normalised by the carrying capacity $C_0$), and red lines show $\bar{A}^*$. Stable branches are indicated by solid lines, and unstable branches by dashed lines. In (b), we set $\Pi_c = 3$. Otherwise, all relevant parameters have the values given in Table \ref{tab3}.
    } \label{bif_Pic_deltac}
\end{figure}

Figure \ref{bif_Pic_deltac} shows additional bifurcation diagrams for the steady states \eqref{SS_triv} and \eqref{SS_non_triv} as functions of $\Pi_c$ (Figure \ref{bif_Pic}) and $\delta_c$ (Figure \ref{bif_deltac}). Since the condition \eqref{cond1} is \emph{always} satisfied for the baseline parameter values, we find that the nontrivial steady state \eqref{SS_non_triv} is stable for all $\Pi_c>0$. Indeed, as $\Pi_c\to\infty$, the phenotypic switching rate is maximised and $C^*/C_0\to1-\frac{\beta_c+\delta_c}{\rho_c}=0.6$ while $\bar{A}^*\to\infty$. Destabilisation of the nontrivial steady state \eqref{SS_non_triv} can be achieved by increasing either $\beta_c$ or $\delta_c$. This occurs trivially if $\beta_c>\rho_c$, but the situation for $\delta_c$ is more complicated. Figure \ref{bif_deltac} shows that, as $\bar{A}^*=1+k$ is a decreasing function of $\delta_c$, the net steady state phenotypic switching rate increases very slowly with increasing $\delta_c$. As such, for the baseline parameter values with $\Pi_c=3$, we find that the stable steady state SMC population $C^*$ remains positive until $\delta_c\approx6350$.    

\section{Results from the full model  \label{rsec2}}
\begin{figure}%[h!]
	\centering
	\begin{subfigure}[b]{0.49\textwidth}
		\centering	
		\includegraphics[height=4.4cm]{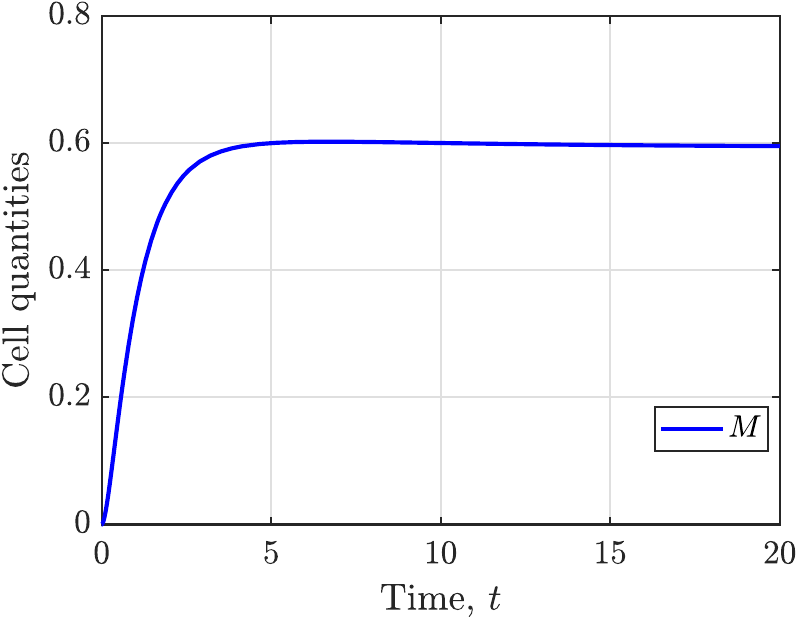}
		\caption{} \label{cells_time_no_SMCs}
	\end{subfigure}
	\begin{subfigure}[b]{0.49\textwidth}
		\centering	
		\includegraphics[height=4.4cm]{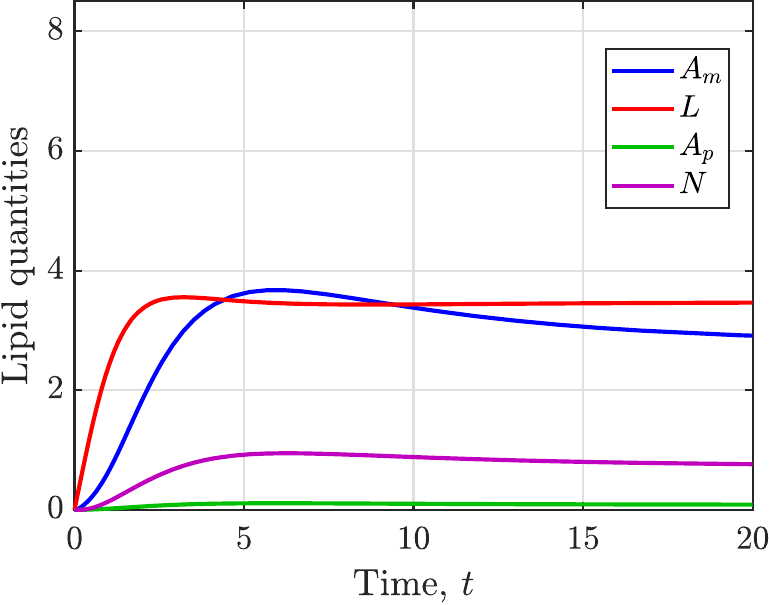}
		\caption{} \label{lipids_time_no_SMCs}
	\end{subfigure}
	\par \medskip
	\begin{subfigure}[b]{0.49\textwidth}
		\centering	
		\includegraphics[height=4.4cm]{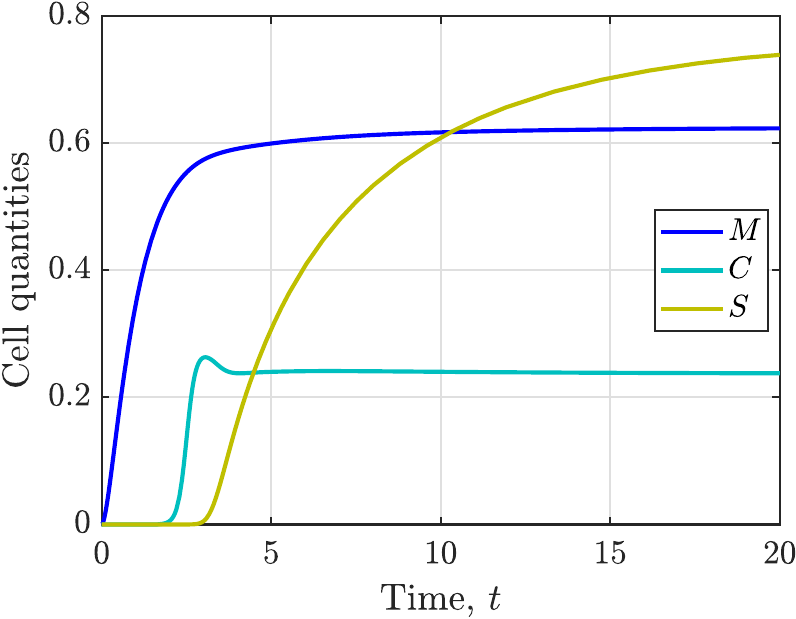}
		\caption{} \label{cells_time_all}
	\end{subfigure}
	\begin{subfigure}[b]{0.49\textwidth}
		\centering	
		\includegraphics[height=4.4cm]{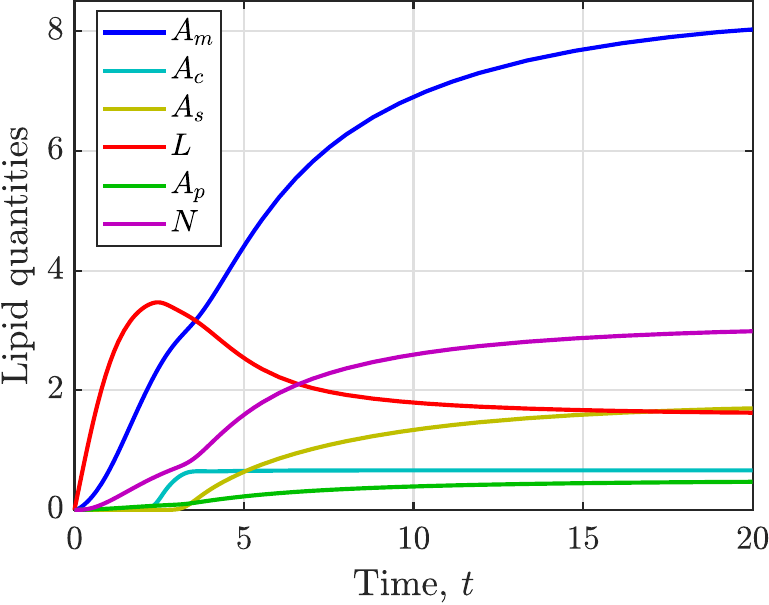}
		\caption{} \label{lipids_time_all}
	\end{subfigure}
    \par \medskip
    \begin{subfigure}[b]{0.49\textwidth}
		\centering	
		\includegraphics[height=4.4cm]{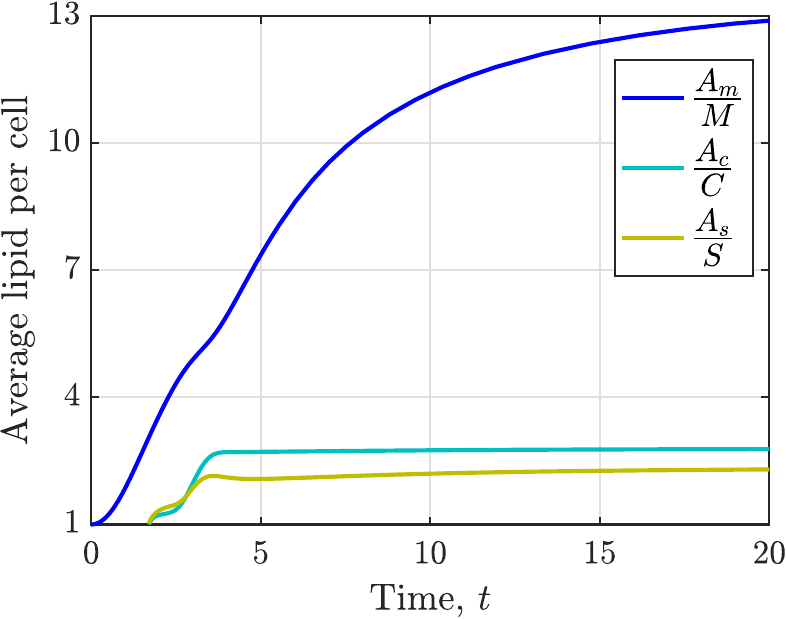}
		\caption{} \label{avg_lipid_time_all}
	\end{subfigure}
	\begin{subfigure}[b]{0.49\textwidth}
		\centering	
		\includegraphics[height=4.4cm]{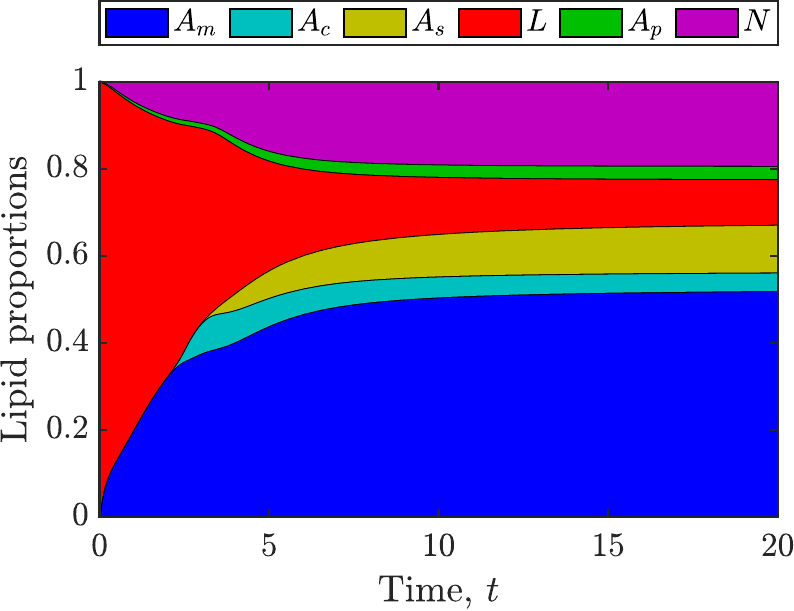}
		\caption{} \label{lipid_props_time_all}
	\end{subfigure}
	\par \medskip
	\begin{subfigure}[b]{0.49\textwidth}
		\centering	
		\includegraphics[height=4.4cm]{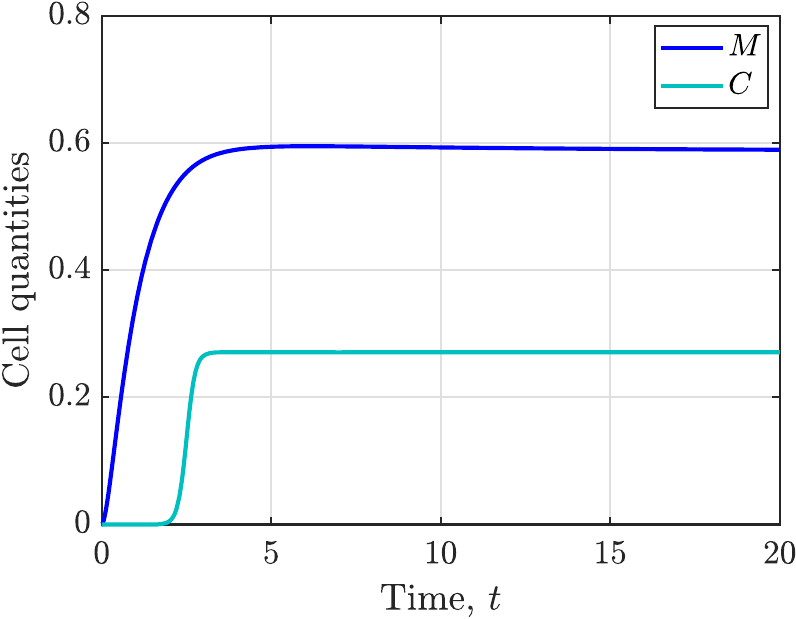}
		\caption{} \label{cells_time_no_SDMs}
	\end{subfigure}
	\begin{subfigure}[b]{0.49\textwidth}
		\centering	
		\includegraphics[height=4.4cm]{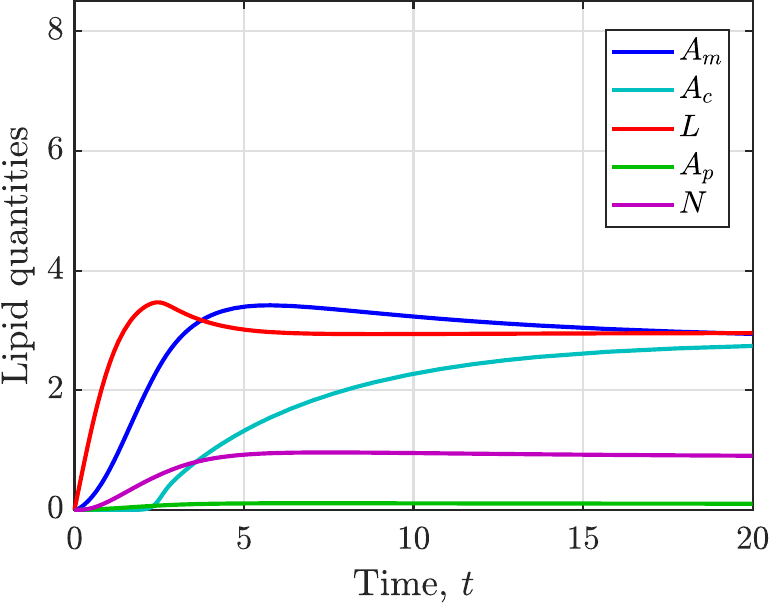}
		\caption{} \label{lipids_time_no_SDMs}
	\end{subfigure}
	\caption{Time-dependent solutions of the model system \eqref{MS}. Panels (a) and (b) show, respectively, the time evolution of the cell and lipid quantities with MDMs only ($C_{init}=0$). Panels (c), (d), (e) and (f), show, respectively, the time evolution of the cell quantities, lipid quantities, average cellular lipid loads, and compartmental lipid proportions with MDMs, SMCs and SDMs all included. Panels (g) and (h) show, respectively, the time evolution of the cell quantities and lipid quantities with MDMs and SMCs only ($\delta_c=0$). Unless otherwise stated, all parameter values are given in Table \ref{tab3}.
    } \label{ODE_sols_time}
\end{figure}

Figure \ref{ODE_sols_time} shows time-dependent solutions for the atherosclerotic plaque cell and lipid dynamics for the complete model system \eqref{MS}. We first set $C_{init}=0$, and allow the MDM population to evolve in isolation (Figures \ref{cells_time_no_SMCs} and \ref{lipids_time_no_SMCs}). The initial accumulation of extracellular modLDL lipid elicits a near-maximal MDM recruitment response, which is then sustained by the subsequent internalisation of modLDL lipid, endogenous lipids, and lipids from apoptotic and necrotic cells. Eventually, both the MDM numbers and the total MDM lipid load reach a peak before decreasing slightly to equilibrium.

Results for the model with MDMs, SMCs, and SDMs are shown in Figures \ref{cells_time_all} to \ref{lipid_props_time_all}. In this simulation, the plaque is seeded with a small initial SMC population $C_{init}=0.0005$ at time $t=t_c=1.7$. The rapid proliferation of these cells quickly drives the SMC population size towards its phenotype switching-free equilibrium (Figure \ref{cells_time_all}). As the per capita SMC proliferation rate declines, the average SMC lipid load increases (Figure \ref{avg_lipid_time_all}), which drives the phenotypic switching of SMCs to SDMs. The sustained phenotypic switching of SMCs to SDMs causes a small reduction in the SMC population and, in the long-term, leads SDMs to become the dominant cell type in the plaque (Figure \ref{cells_time_all}).

Comparing Figure \ref{lipids_time_all} with Figure \ref{lipids_time_no_SMCs} shows that the inclusion of SMCs and SDMs in the model leads to a greater than 2-fold increase in the total lipid held in the plaque. As the rate of modLDL influx remains constant, and the MDM dynamics are largely unaffected, we conclude that this increase is caused by the substantial \emph{de novo} synthesis of endogenous lipid by the rapidly proliferating SMCs and SDMs. Naturally, some of this additional lipid remains internalised in the SMCs and SDMs. However, we also observe increases in the long-term quantities of MDM lipid, necrotic lipid, and apoptotic lipid (around 3-, 4-, and 5-fold greater than in Figure \ref{lipids_time_no_SMCs}, respectively). Factors that most likely contribute to these increases are: (1) the absence of SDM (and SMC) emigration out of the plaque, and (2) the relatively large rate of SDM apoptosis. Noticeably, long-term modLDL lipid levels are reduced by around 50\% because there are now more cells in the plaque, including SMCs, that can internalise lipid from modLDL.

Figure \ref{lipid_props_time_all} shows how the proportion of lipid in each model compartment changes over time. Although the proportion of lipid in modLDL is initially decreasing, and the proportion of lipid in MDMs, apoptotic cells and necrotic cells is initially increasing, a noticeable change in the dynamics occurs upon the emergence of SDMs at around $t=3$. As the proportion of lipid in the SDMs grows, there is an acceleration in the growth of the proportions of necrotic, apoptotic, and MDM lipid at the expense of the proportions in SDMs and modLDL.

\begin{figure}%[h!]
	\centering
\includegraphics[height=6cm]{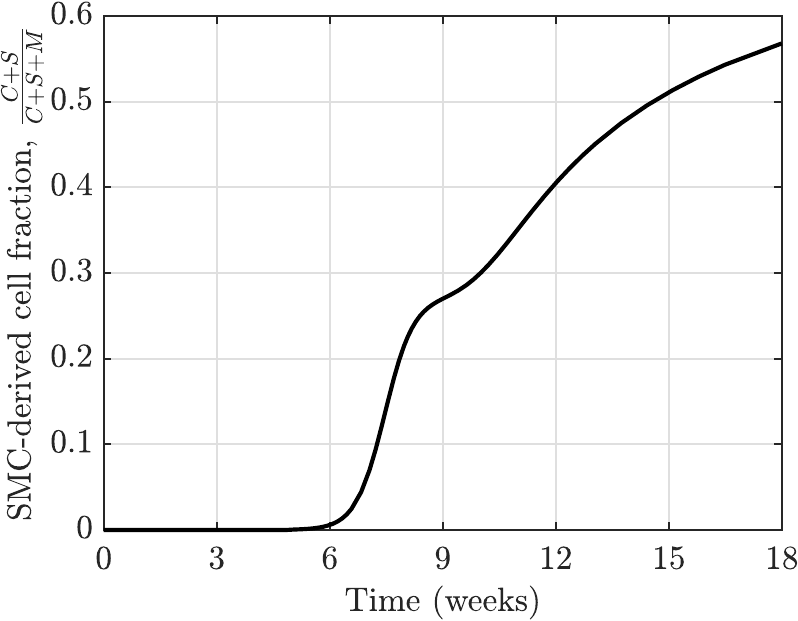}
	\caption{Plot showing the time evolution of the fraction of cells in the model plaque that are derived from the initial SMC population. The curve is plotted against physical time to admit comparison with the experimental data in Figure 2d of \cite{misra2018integrin}. 
    }\label{SMC_fract}
\end{figure}

Figure \ref{SMC_fract} plots the fraction of SMC-derived cells in the model plaque, $\frac{C+S}{C+S+M}$, against physical time. Consistent with observations from lineage tracing experiments in mouse models \cite{misra2018integrin,shankman2015klf4}, this plot demonstrates that cells of SMC origin eventually outweigh the proportion of MDMs in the plaque. The temporal evolution of the SMC-derived cell fraction predicted by the model compares favourably with the experimental measurements at weeks 6, 12, and 16 shown in Figure 2b of \cite{misra2018integrin}.

To verify that the increased amounts of apoptotic, necrotic, and MDM lipid seen in Figure \ref{lipids_time_all} are indeed due to the presence of SDMs, we set $\delta_c=0$ to simulate a scenario where SMCs cannot change phenotype (Figures \ref{cells_time_no_SDMs} and \ref{lipids_time_no_SDMs}). In this case, we observe relatively minor increases in the apoptotic, necrotic and MDM lipid. This confirms that it is the emergence of SDMs via phenotype switching that drives the previously observed increase in total lipid in these compartments. Comparing Figures \ref{lipids_time_no_SDMs} and \ref{lipids_time_all} shows that, in the absence of phenotype switching, the SMC population holds around 4 times more lipid in the long-term. However, in this case, there are no detrimental effects that arise from this increased lipid loading.

\begin{figure}[h!]
	\centering
	\begin{subfigure}[b]{0.49\textwidth}
		\centering
		\includegraphics[height=5cm]{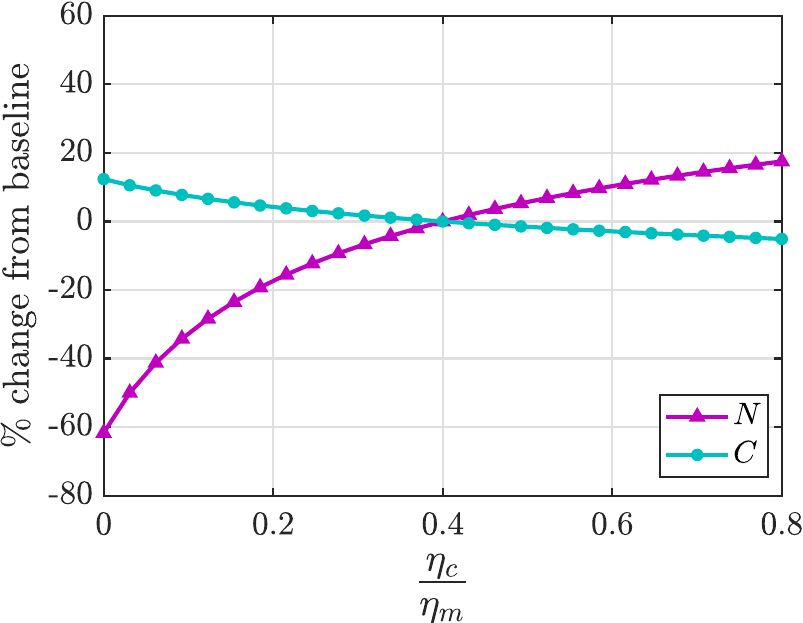}
        \caption{} \label{N_C_sens_eta}
	\end{subfigure}
	\begin{subfigure}[b]{0.49\textwidth}
		\centering
		\includegraphics[height=5cm]{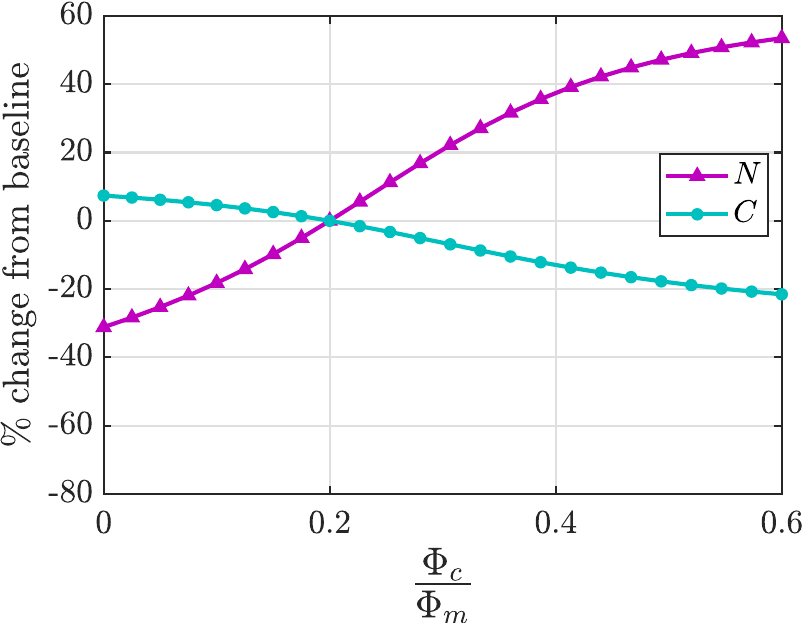}
        \caption{} \label{N_C_sens_Phi}
	\end{subfigure}
	\par \medskip
	\begin{subfigure}[b]{0.49\textwidth}
		\centering
		\includegraphics[height=5cm]{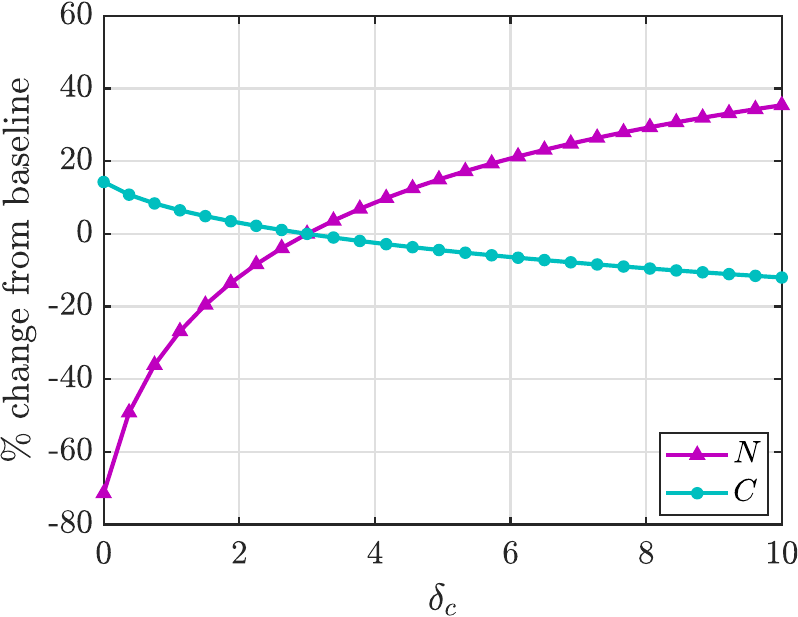}
        \caption{} \label{N_C_sens_delta}
	\end{subfigure}
	\begin{subfigure}[b]{0.49\textwidth}
		\centering
		\includegraphics[height=5cm]{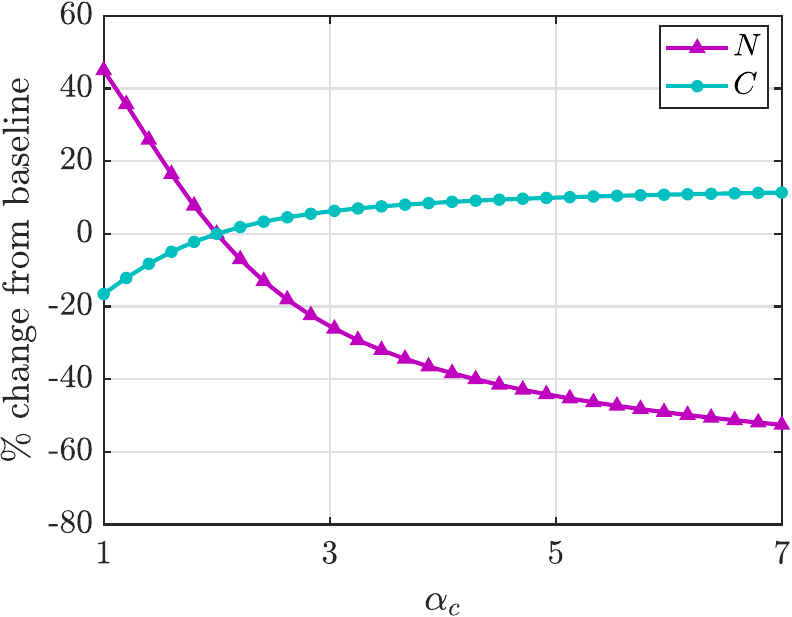}
        \caption{} \label{N_C_sens_alpha}
	\end{subfigure}	
	\caption{Plots showing the percentage change in steady state SMC numbers (cyan lines) and necrotic lipid mass (magenta lines) in the model plaque as the parameters (a) $\eta_c$, (b) $\Phi_c$, (c) $\delta_c$, and (d) $\alpha_c$ are independently varied from their baseline values.}\label{N_C_sens}
\end{figure}

\begin{figure}[h!]
	\centering
	\begin{subfigure}[b]{0.49\textwidth}
		\centering
		\includegraphics[height=5cm]{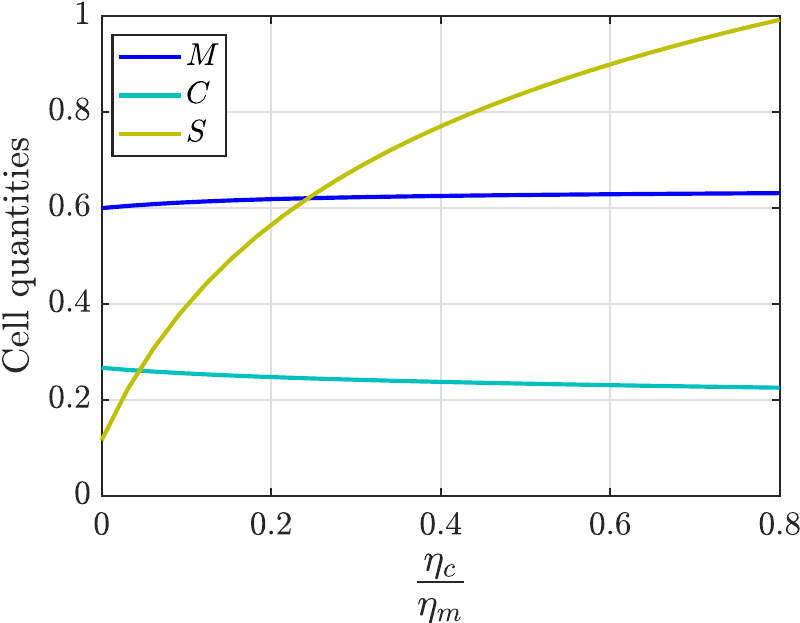}
        \caption{} \label{cells_sens_etac}
	\end{subfigure}
	\begin{subfigure}[b]{0.49\textwidth}
		\centering
		\includegraphics[height=5cm]{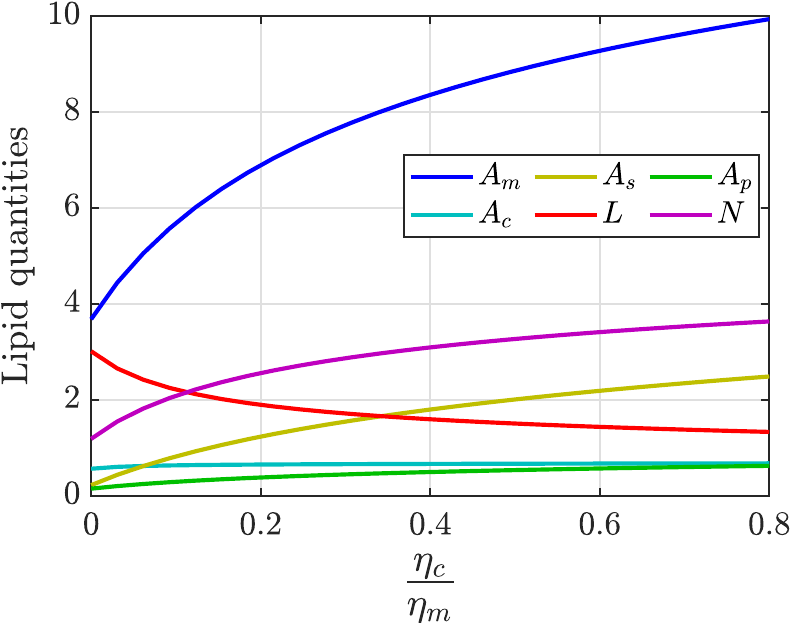}
        \caption{} \label{lipids_sens_etac}
	\end{subfigure}
	\par \medskip
	\begin{subfigure}[b]{0.49\textwidth}
		\centering
		\includegraphics[height=5cm]{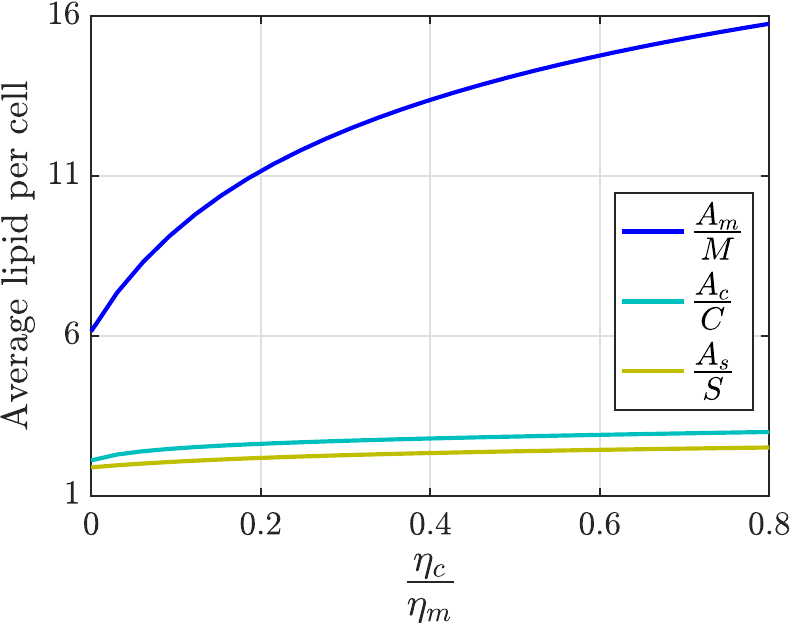}
        \caption{} \label{avg_lipid_sens_etac}
	\end{subfigure}
	\begin{subfigure}[b]{0.49\textwidth}
		\centering
		\includegraphics[height=5cm]{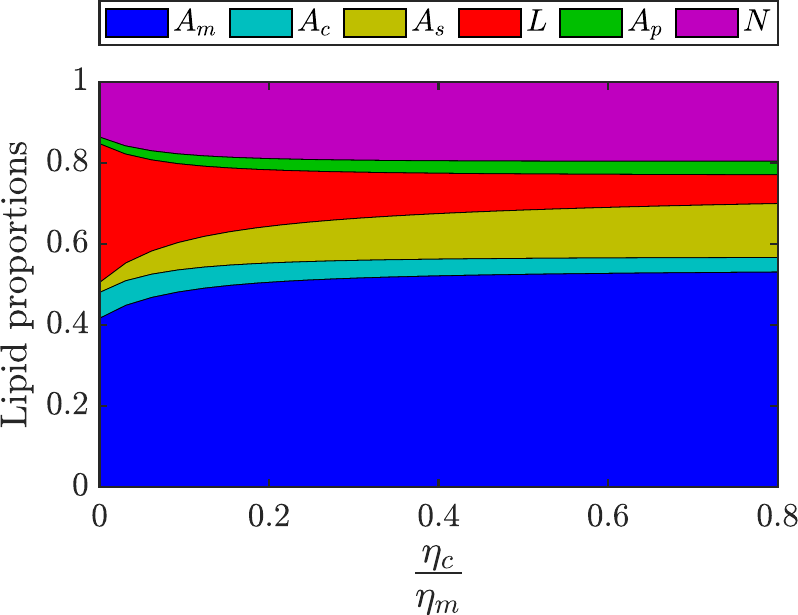}
        \caption{} \label{lipid_props_SS_sens_etac}
	\end{subfigure}
	\par \medskip
	\begin{subfigure}[b]{\textwidth}
		\centering
		\includegraphics[height=7cm]{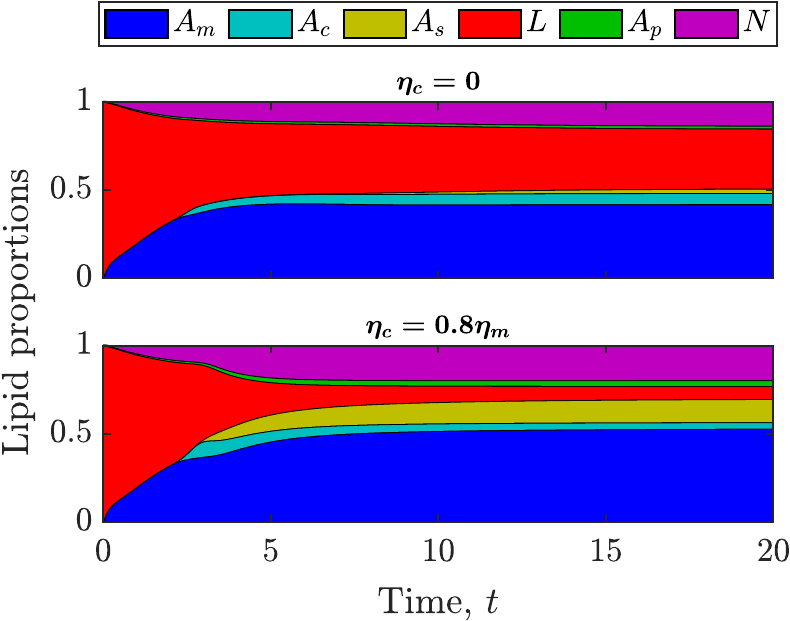}
        \caption{} \label{lipid_props_time_sens_etac}
	\end{subfigure}
	\caption{Effect of varying the rate of SMC modLDL consumption over the range $\eta_c\in[0,0.8\eta_m]$. Panels show steady state (a) cell quantities, (b) lipid quantities, (c) average cellular lipid loads, and (d) compartmental lipid proportions as a function of $\eta_c$. Panel (e) shows the time evolution of compartmental lipid proportions for $\eta_c=0$ (top) and $\eta_c=0.8\eta_m$ (bottom). For all simulations, the MDM modLDL consumption rate $\eta_m$ is held at its baseline value.} \label{sens_etac}
\end{figure}

Figure \ref{N_C_sens} presents plots that show the percentage change in steady state necrotic lipid and cap SMCs as parameters that control SMC lipid consumption ($\eta_c$, $\Phi_c$) and phenotypic switching ($\delta_c$, $\alpha_c$) are varied in the model. Long-term SMC numbers and necrotic lipid mass are important indicators of the clinical risk associated with a plaque because cap SMCs provide plaque stability, and the release of necrotic lipid from a ruptured plaque promotes dangerous blood clot formation. Figures \ref{cells_sens_etac} to \ref{lipid_props_SS_sens_etac} present complementary information for the case in which $\eta_c$ is varied (Figure \ref{N_C_sens_eta}) by showing details of corresponding changes in cell numbers, lipid quantities, average cellular lipid loads, and compartmental lipid proportions.

In Figures \ref{N_C_sens_eta} and \ref{cells_sens_etac} to \ref{lipid_props_SS_sens_etac}, the horizontal axis shows the rate of consumption of modLDL by SMCs ($\eta_c$) relative to that by MDMs ($\eta_m$). We observe that as $\eta_c$ increases from $0$ to $0.8\eta_m$, SMC numbers reduce by around 15\%, while the SDM population increases almost 9-fold (Figure \ref{cells_sens_etac}). This is because the phenotypic switch that SMCs undergo is driven by lipid loading, so that if SMCs ingest lipid more rapidly, they are more likely to adopt a macrophage-like phenotype. The MDM population is relatively insensitive to changes in modLDL uptake by SMCs. Thus, as $\eta_c$ increases beyond about $0.25\eta_m$, SDMs become the dominant cell type in the plaque at steady state (Figure \ref{cells_sens_etac}).

Figure \ref{lipids_sens_etac} indicates that, for all $\eta_c\in[0,0.8\eta_m]$, the MDM population is the compartment that carries the most lipid. This lipid load increases around 2.5-fold with $\eta_c$, reflecting an increase in the average cellular lipid load from around 6 to 16 lipid units (Figure \ref{avg_lipid_sens_etac}). The amount of necrotic lipid at steady state increases by a similar extent to the MDM lipid over the range of $\eta_c$ considered. Predictably, as $\eta_c$ increases, a lower proportion of lipid is in modLDL and a higher proportion is in inside SDMs (Figures \ref{lipid_props_SS_sens_etac} and \ref{lipid_props_time_sens_etac}). The proportion of lipid in the other compartments does not change significantly at steady state (Figure \ref{lipid_props_SS_sens_etac}), but it must be remembered that there is much more lipid in the system overall when $\eta_c$ is large.

The results in Figures \ref{lipid_props_SS_sens_etac} and \ref{lipid_props_time_sens_etac} indicate that, for small $\eta_c$, MDMs are the major lipid-handling cells in the plaque. However, as $\eta_c$ increases, SDMs bear an increasing lipid burden. Since, in this model, SDMs cannot emigrate out of the plaque, and have limited capacity for lipid export to HDL, increasing lipid uptake by SDMs results in MDMs accumulating more lipid via phagocytosis of apoptotic and necrotic lipid contributed by SDM death. This suggests that lipid consumption by SDMs may be linked to a higher risk of unstable, lipid-filled plaques that can lead to clinical sequelae such as heart attack and stroke.

Figure \ref{N_C_sens_Phi} shows the percentage change in steady state SMC numbers and necrotic lipid mass relative to baseline ($\Phi_c=0.2\Phi_m$) as the SMC efferocytosis rate is varied over the range $\Phi_c\in[0,0.6\Phi_m]$. The results are qualitatively similar to those for the case where $\eta_c$ is varied (Figure \ref{N_C_sens_eta}), showing an increase in necrotic lipid accumulation and a decrease in cap SMC numbers as $\Phi_c$ increases. Figures \ref{N_C_sens_eta} and \ref{N_C_sens_Phi} therefore confirm that the mechanism of SMC phenotypic switching assumed in this model is independent of the origin of the lipid consumed by SMCs.

The results in Figures \ref{N_C_sens_delta} and \ref{N_C_sens_alpha} illustrate the impact of varying the maximum SMC phenotypic switching rate $\delta_c$, and the average ingested lipid load for half-maximal SMC switching $\alpha_c$, respectively, on SMC numbers and necrotic lipid at steady state. Consistent with the observations in Figure \ref{ODE_sols_time}, reducing $\delta_c$ from its baseline value towards zero results in a dramatic reduction in necrotic lipid and a mild increase in SMC numbers. Increasing $\delta_c$ above its baseline value reverses these trends, but the necrotic lipid quantity is less sensitive to changes in $\delta_c$ in this case (Figure \ref{N_C_sens_delta}). For $\alpha_c$, we observe a trend of increasing SMC numbers and decreasing necrotic lipid as $\alpha_c$ is increased (Figure \ref{N_C_sens_alpha}). This is because increasing $\alpha_c$ suppresses the net phenotypic switching rate of SMCs to SDMs when the average ingested SMC lipid load is small. In practice, this can lead to a temporal delay in the emergence of a SDM population. For example, with $\alpha_c=7$, we find that SDM numbers remain at negligible levels ($<C_{init}$) for approximately 2 weeks of physical time longer than for the baseline case (results not shown).

\begin{figure}[h!]
	\centering
	\begin{subfigure}[b]{0.49\textwidth}
		\centering
		\includegraphics[height=5cm]{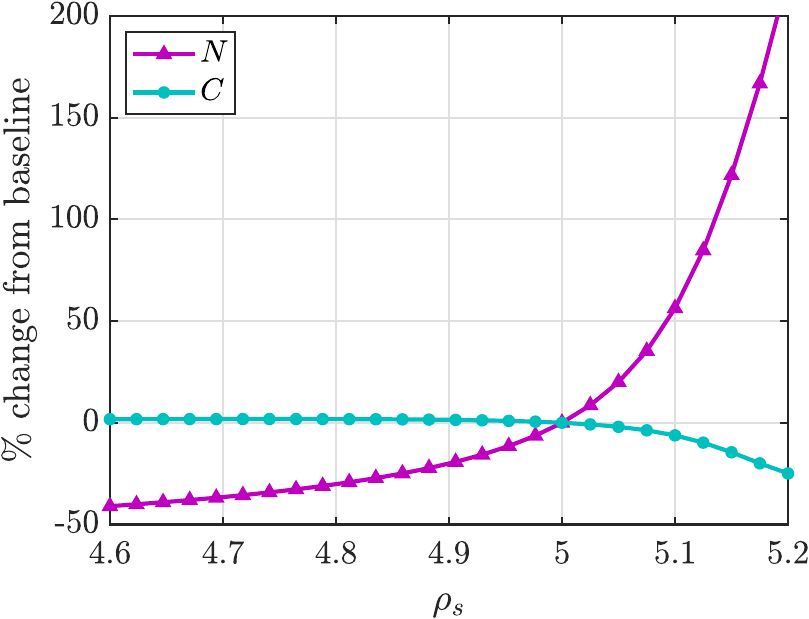}
        \caption{}\label{N_C_sens_rho}
	\end{subfigure}
	\begin{subfigure}[b]{0.49\textwidth}
		\centering
		\includegraphics[height=5cm]{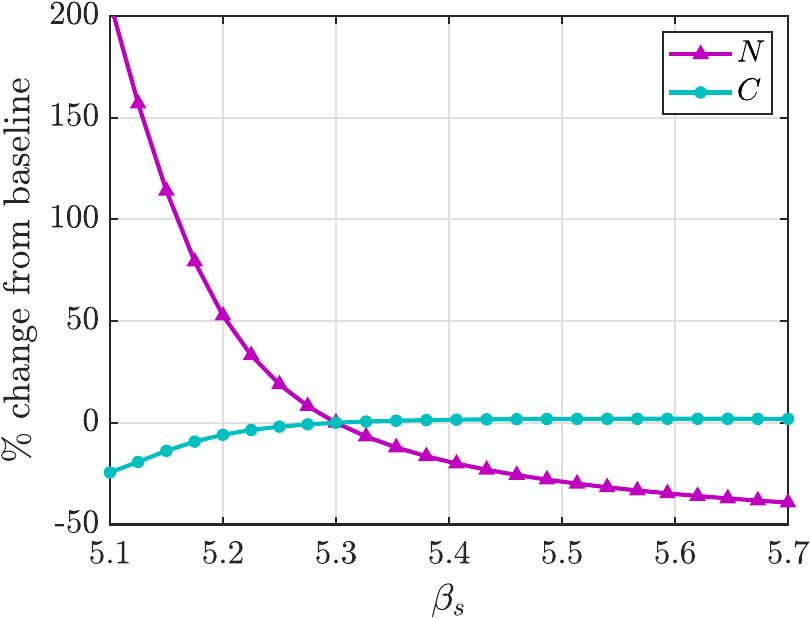}
        \caption{}\label{N_C_sens_beta}
	\end{subfigure}
	\caption{Plots showing the percentage change in steady state SMC numbers (cyan lines) and necrotic lipid mass (magenta lines) in the model plaque as the parameters (a) $\rho_s$, and (b) $\beta_s$ are independently varied from their baseline values.}\label{sens_rho_beta}
\end{figure}

\begin{figure}[h!]
	\centering
	\includegraphics[height=5cm]{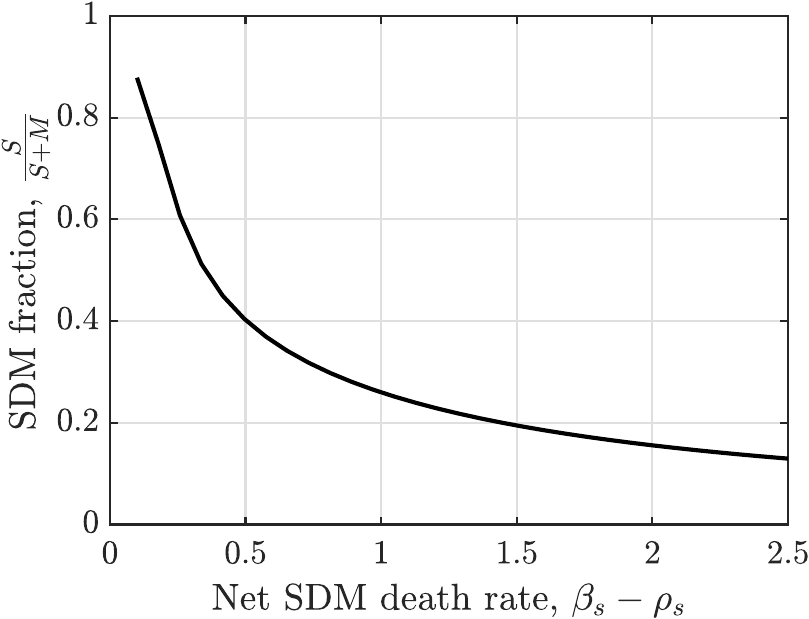}
	\caption{Plot showing the steady state fraction of macrophages that are SDMs as a function of the net SDM death rate, $\beta_s-\rho_s$. For this plot, $\beta_s$ is held at its baseline value, while $\rho_s$ is varied over the range $[2.8,5.2]$.}\label{SDM_fract_vs_death}
\end{figure}

\begin{figure}[h!]
	\centering
	\begin{subfigure}[b]{0.49\textwidth}
		\centering
		\includegraphics[height=5cm]{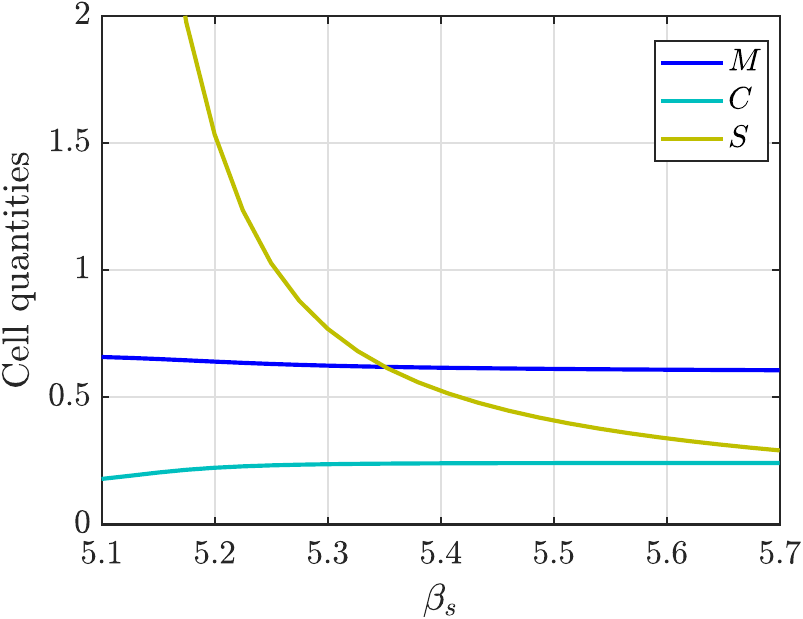}
        \caption{}\label{cells_sens_betaS}
	\end{subfigure}
	\begin{subfigure}[b]{0.49\textwidth}
		\centering
		\includegraphics[height=5cm]{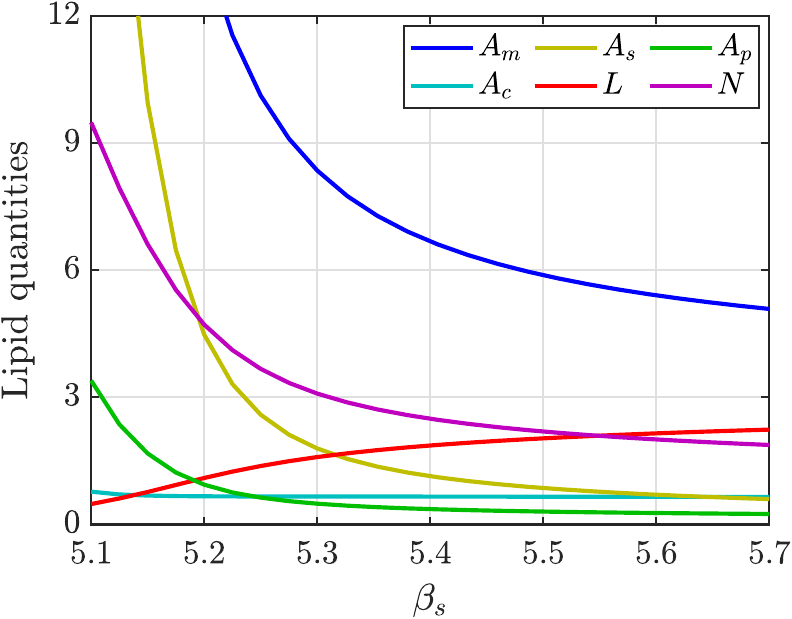}
        \caption{}\label{lipid_sens_betaS}
	\end{subfigure}
	\par \medskip
	\begin{subfigure}[b]{0.49\textwidth}
		\centering
		\includegraphics[height=5cm]{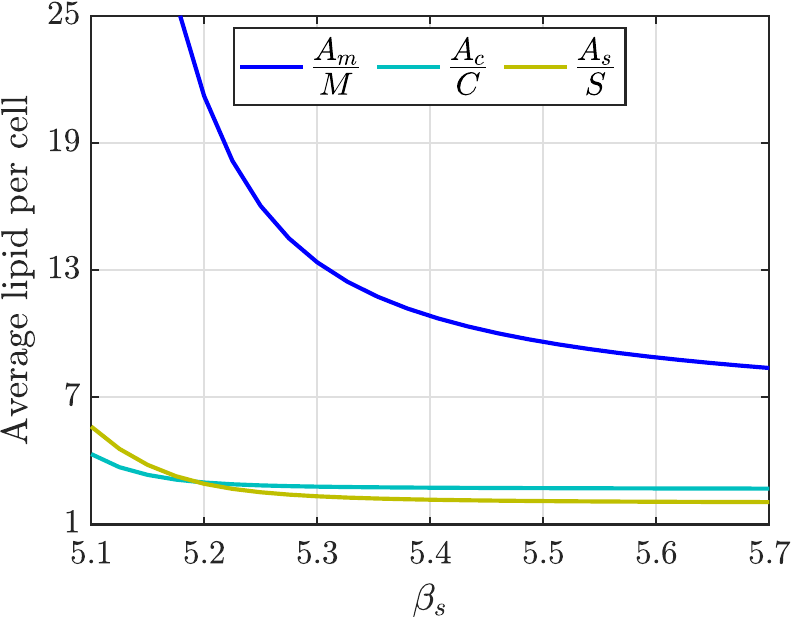}
        \caption{}\label{avg_lipid_sens_betaS}
	\end{subfigure}
	\begin{subfigure}[b]{0.49\textwidth}
		\centering
		\includegraphics[height=5cm]{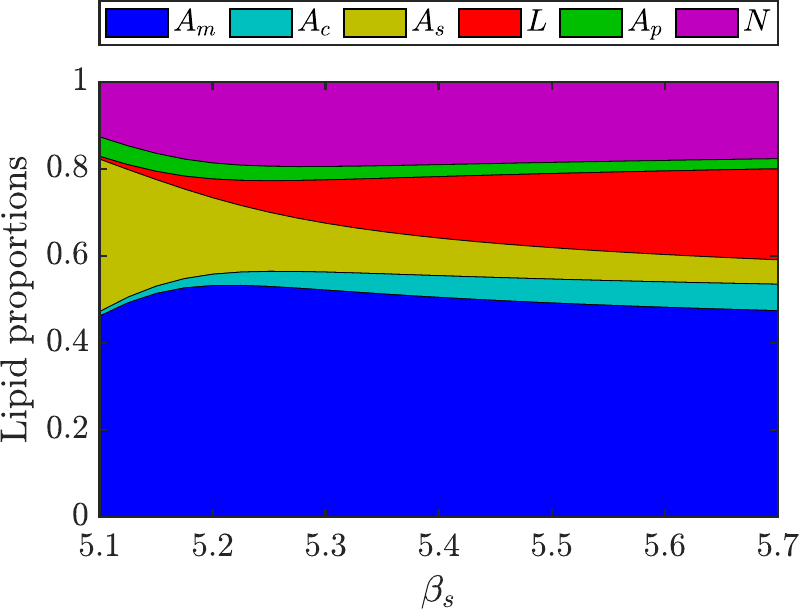}
        \caption{}\label{lipid_props_SS_sens_betaS}
	\end{subfigure}
	\caption{Effect of varying the SDM apoptosis rate over the range $\beta_s\in[5.1,5.7]$. Panels show steady state (a) cell quantities, (b) lipid quantities, (c) average cellular lipid loads, and (d) compartmental lipid proportions as a function of $\beta_s$.}\label{sens_betas}
\end{figure}

Figures \ref{N_C_sens_rho} and \ref{N_C_sens_beta} illustrate the percentage changes in steady state SMC numbers and necrotic lipid as the SDM proliferation rate $\rho_s$, and the SDM apoptosis rate $\beta_s$ are independently varied from their baseline values. To ensure that the SDM population does not grow without bound, we consider only values of $\rho_s$ and $\beta_s$ for which the net SDM death rate is positive (i.e., $\beta_s-\rho_s>0$). As $\rho_s$ increases above baseline, or $\beta_s$ decreases below baseline, the steady state necrotic lipid increases dramatically. This is due to a substantial increase in the steady state SDM population (see Figure \ref{cells_sens_betaS}). The fact that Figures \ref{N_C_sens_rho} and \ref{N_C_sens_beta} are almost mirror images implies that the net SDM death rate $\beta_s-\rho_s$ in equation \eqref{Seqn} is the key parameter combination that underlies these results. This is despite the fact that $\rho_s$ and $\beta_s$ appear independently in equation \eqref{Aseqn}. As the net SDM death rate approaches zero, there is a rapid increase in the proportion of plaque macrophages, at steady state, that are SDMs (Figure \ref{SDM_fract_vs_death}). This corresponds to a highly pathological and dangerous plaque state.

A more detailed illustration of the effects of changing the SDM apoptosis rate $\beta_s$ (for fixed $\rho_s$) is given in Figure \ref{sens_betas}. For $\beta_s$ close to $\rho_s$ (small net death rate), the dramatic increase in steady state SDM numbers is accompanied by a dramatic increase in the amount of lipid held in both the SDM and MDM populations (Figure \ref{lipid_sens_betaS}). However, the average lipid per cell in MDMs at steady state is much higher than the average lipid per cell in SDMs (Figure \ref{avg_lipid_sens_betaS}), and MDMs carry a larger proportion of the plaque intracellular lipid (Figure \ref{lipid_props_SS_sens_betaS}). As well as presumably contributing to substantial plaque growth, our results suggest that a small net SDM death rate may lead to increased plaque inflammation as heavily lipid-laden MDMs are likely to release more inflammatory cytokines.

\begin{figure}[h!]
	\centering
		\includegraphics[width=0.850\linewidth]{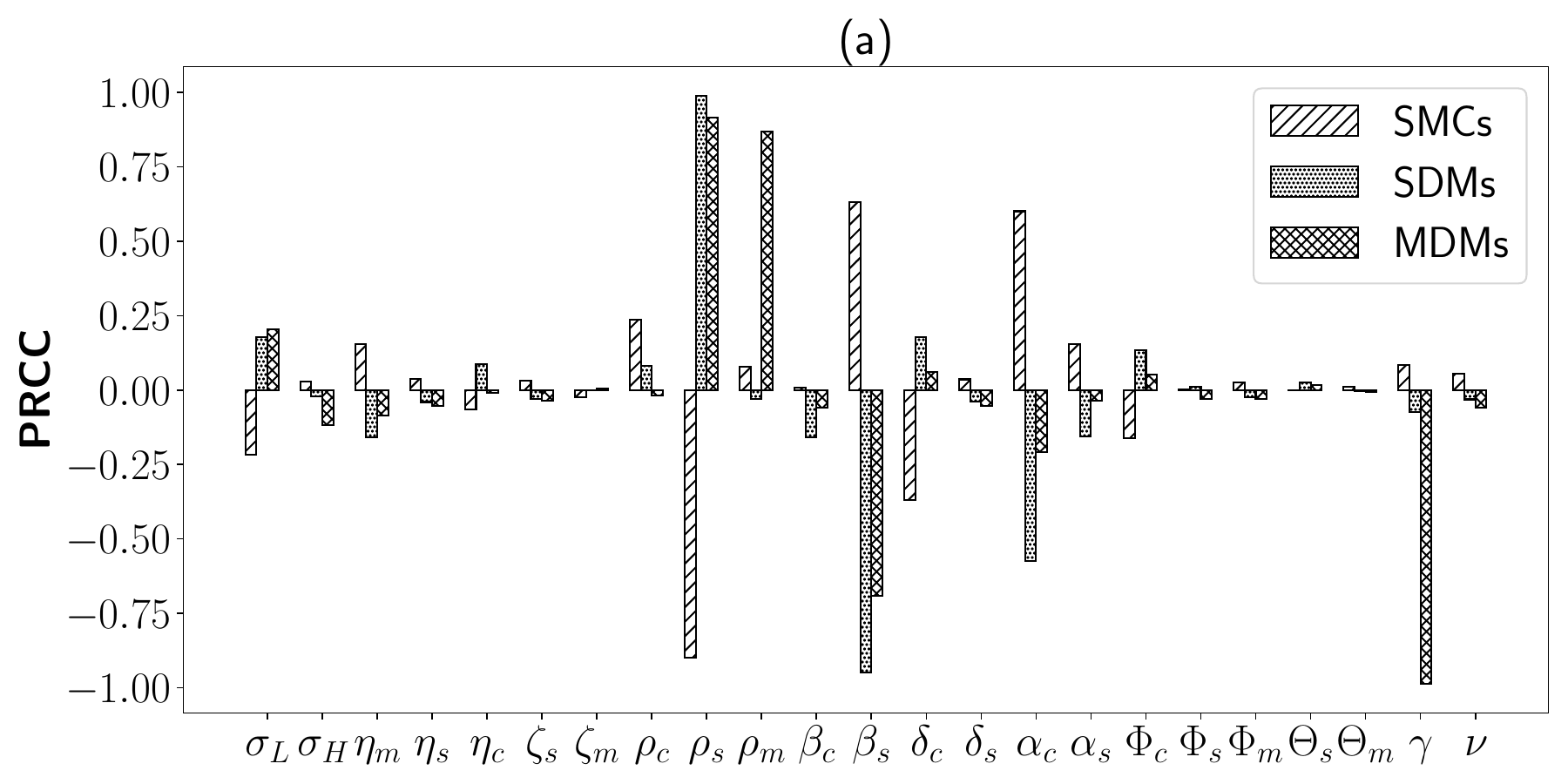}
        \includegraphics[width=0.850\linewidth]{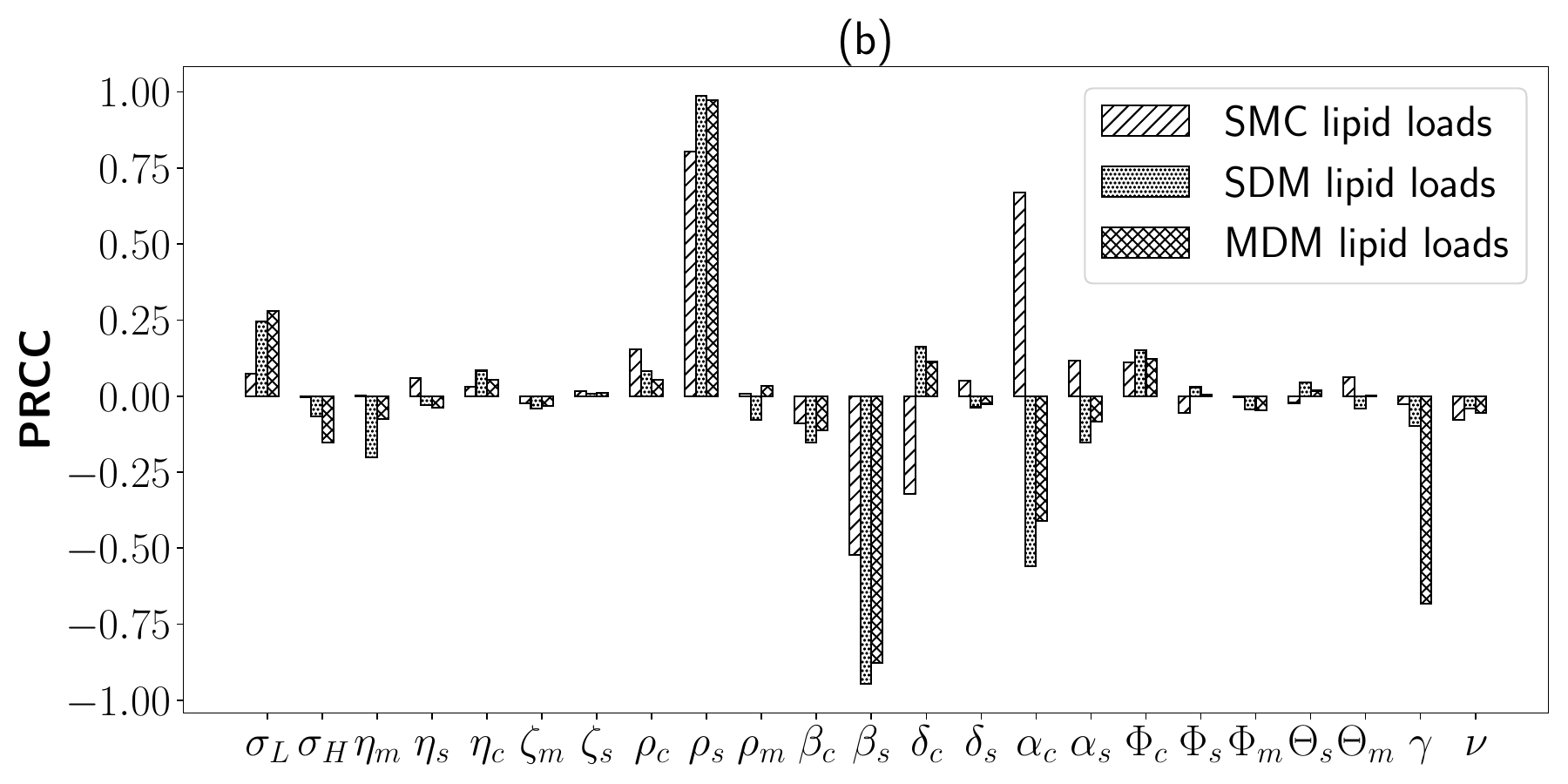}
        \includegraphics[width=0.850\linewidth]{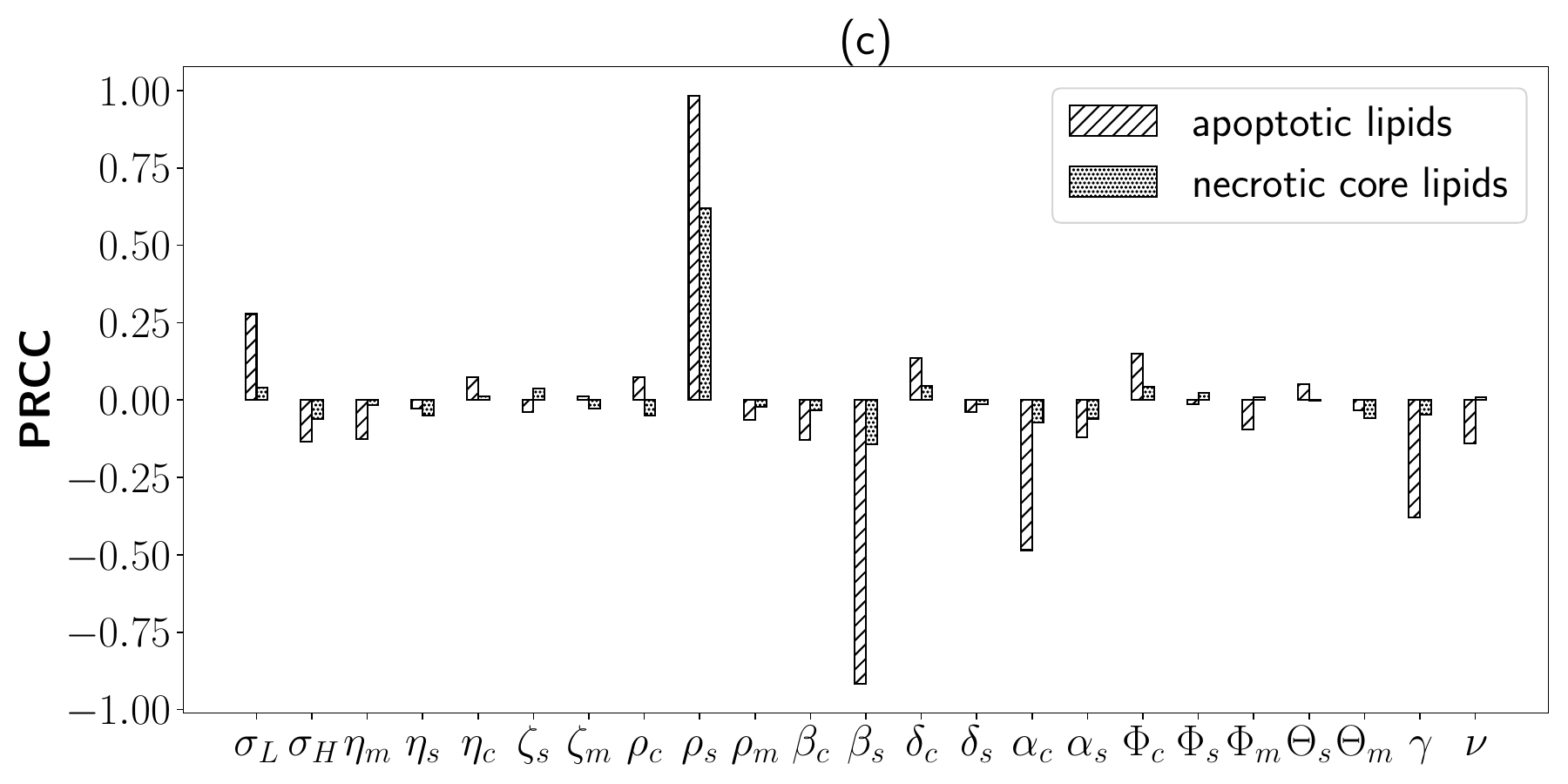}
        \caption{Results of a parameter sensitivity analysis on the model \eqref{MS} showing PRCC values for each model parameter with respect to steady state (a) cell quantities, (b) intracellular lipid quantities, and (c) apoptotic and necrotic lipid quantities. Parameters were varied by $\pm20\%$ from their baseline values, and the baseline parameter values used in this analysis are listed in Table  \ref{tab3}. Samples violating the constraint $\rho_s\le\beta_s$ were excluded from the analysis to avoid scenarios with unrealistic growth of the SDM population.}\label{SA}
\end{figure}

To comprehensively study the effect of varying parameter values away from baseline on the steady state solutions of the model, we used Latin hypercube sampling and calculated partial rank correlation coefficients (PRCCs) \cite{blower1994sensitivity,marino2008methodology} with 1000 simulations per run. The results of this sensitivity analysis are shown in Figure \ref{SA}. For a given model variable and parameter, a PRCC value close to $+1$ or $-1$ indicates a strong positive or negative correlation, respectively, between the parameter and the value of the variable at steady state. Positive PRCC values indicate that the value of the variable increases as the value of the parameter increases, while negative PRCC values indicate that the value of the variable decreases as the value of the parameter increases. Figure \ref{SA} shows that the MDM emigration rate, and the MDM and SDM proliferation and apoptosis rates ($\gamma$, $\rho_m$, $\rho_s$, $\beta_s$) clearly have the most impact on cell populations. (Recall that the MDM apoptosis rate $\beta_m$ has been scaled to $1$ in the dimensionless system \eqref{MS}, so is not explicitly varied in this analysis). The values $\rho_s$ and $\beta_s$ also have a significant impact on the intracellular lipids, and the apoptotic and necrotic lipids, as shown above. Interestingly, the parameter related to phenotypic switching that has the greatest impact is $\alpha_c$, which defines the average ingested lipid load for a half-maximal SMC switching rate. In addition to the expected influence on steady state SMC and SDM populations, $\alpha_c$ also strongly impacts the lipid held in each of the three cell types as well as in apoptotic cells.  

\section{Discussion} \label{s4}
This paper presents an ODE model of SMC phenotypic switching in the atherosclerotic plaque. We use the model to explore the impact of SMC phenotypic switching on plaque progression, with a focus on cell dynamics and lipid accumulation. The model assumes that plaque cells can ingest lipid from modified LDL particles and dead cells, and that plaque cells can efflux lipid to HDL particles. The model also includes recruitment and emigration of MDMs, and proliferation and death of all plaque cell types. We explore how SMC phenotypic switching into macrophage-like cells (SDMs) can contribute to plaque instability by increasing necrotic core lipid and depleting the fibrous cap.

The modelling assumptions incorporate several significant observations from recent experimental studies. For example, the murine plaque lineage tracing study by Misra et al. \cite{misra2018integrin} showed that the plaque SMC population is initiated by one or two highly proliferative progenitor cells that enter the plaque after 5--6 weeks of feeding on a high-fat diet. This is modelled by assuming that the SMC population is zero for the first 5 weeks of physical time before a very small number of rapidly proliferating SMCs is added to the plaque to initiate population growth. The \emph{in vitro} study by Vengrenyuk et al.\ \cite{vengrenyuk2015cholesterol} showed that cholesterol loading of vascular SMCs leads to reduced expression of typical SMC markers and increased expression of typical macrophage markers. Vengrenyuk et al.\ \cite{vengrenyuk2015cholesterol} further showed that these trends of marker expression can be reversed if SMCs can offload their internalised lipid. These observations are modelled by assuming bidirectional phenotypic switching between SMCs and SDMs, where the respective switching rates are given by appropriately defined functions of internalised cellular lipid loads.

The model results collectively highlight the critical role of SMCs in lipid trafficking within the plaque microenvironment. When SMCs ingest sufficient lipid from modLDL or apoptotic cells, they transition to SDMs, which ultimately contribute to the expansion of the plaque necrotic core. Figure \ref{lipids_time_all}, for example, shows a 4-fold increase in necrotic lipid for the parameter values in Table \ref{tab3}. The loss of functional SMCs from the cap due to phenotypic switching also reduces the cap SMC population, which then relies upon rapid and sustained SMC proliferation to prevent excessive depletion of these critical cells. The combination of necrotic core growth and cap SMC loss that we observe in the presence of SMC phenotypic switching is indicative of heightened plaque vulnerability. The model therefore suggests that inhibiting SMC-to-SDM phenotypic switching, by targeting processes such as SMC lipid consumption, could help to reduce plaque vulnerability and lower cardiovascular risk.

The model leads us to the following conclusions about the impact on plaque fate of SMCs switching phenotype to become macrophage-like SDMs.

\begin{itemize}
 \item The total number of plaque cells with macrophage phenotype can increase significantly when SMCs switch to SDMs. When the net SDM death rate is small, this can result in dramatically more macrophage-like cells in the plaque. This result is in agreement with results from experiments in mice, where observations show that large and growing plaques may contain significant populations of SDMs \cite{Pan_etal_2024}. %When SDM death and proliferation rates are comparable to MDM rates of death and proliferation then the model predicts that the proportion of phenotypically macrophage-like cells that are SDMs is 40\% or greater.  This is in agreement with experimental observations \cite{shankman2015klf4}.
 
 \item In the model, MDM numbers are not significantly altered in the presence of SDMs. However, the average MDM lipid load is significantly increased in the presence of SDMs. Several factors contribute to this phenomenon. First, the rapid and unfettered proliferation of SDMs increases the overall plaque lipid content via the \emph{de novo} endogenous lipid synthesis required to form new daughter cells. Second, it is assumed that SDMs cannot offload internalised lipid to HDL as efficiently as MDMs \cite{Francis_2023}, nor emigrate out of the plaque taking their internalised lipid with them. Hence, the only fate available to SDMs is to die in the plaque. When an SDM dies, its internalised lipid is added to the apoptotic lipid pool. This lipid is mainly efferocytosed by MDMs, whose efferocytosis rate in the model is four times that of SDMs and five times that of SMCs.
 
 \item The average SDM lipid load in the model is considerably smaller than the average lipid load of MDMs. This observation reflects differences between the population dynamics and lipid trafficking properties of the two cell species. SDMs have substantially higher proliferation and death rates compared to MDMs. This means that the lipid in SDMs is frequently divided between daughter cells that ultimately have short lifespans with limited time for lipid consumption. Moreover, SDMs consume all types of lipid at a slower rate than MDMs. Hence, even though the SDM population may be large, a very significant proportion of plaque internalised lipid is still held by MDMs. If our assumptions are valid, we would expect that most of the heavily lipid laden cells in a plaque are MDMs and not SDMs.
 
 \item The growth of the SDM population is driven, in part, by the bounded proliferation of cap SMCs. As SMCs switch phenotype and leave this population, they are replaced by the proliferation of remaining SMCs. If the SMCs do not proliferate or proliferate too slowly, then, in time, the entire SMC population may be lost. In this case, SDM proliferation would be the only possible source of new SDMs.
\end{itemize}

The model reported in this work has some limitations to be addressed in future studies. For example, the ODE formulation of the model means that the phenotypic switching rates are expressed in terms of the average ingested lipid loads across the entire SMC and SDM populations. In reality, the cells within these populations will have a spectrum of ingested lipid loads, and only a particular subset of cells would be likely to undergo phenotype change at any given time. This limitation could be addressed by formulating the model with resolution in cellular lipid loads \cite{ford2019lipid,Chambers_etal_proliferation_2024}. In such a framework, the lipid-dependent cell behaviour could also be extended to include MDMs. Here, we assume that MDMs can acquire very large lipid loads without any loss of normal function. If MDM function were assumed to be impaired by lipid loading \cite{CHAMBERS2023108971,watson2023}, it is possible that the implications of SMC phenotypic switching for plaque fate could be considerably worse than predicted in this study.

A further interesting target for future studies is to formulate the model with resolution in space. This could lead to a more comprehensive understanding of exactly how cap SMCs are exposed to lipid in the plaque, and how the movement of plaque cells (e.g., SDMs vacating the cap region) influence the long-term fate and spatial structure of the plaque. An intriguing possibility would be to combine spatial structure with lipid structure, as reported in a recent model for macrophages in the early human plaque \cite{CHAMBERS2025112232}.

The model proposed in this paper is informed by observations of plaque SMC behaviour in murine models of atherosclerosis, and we therefore urge some caution in interpreting our findings with respect to the human condition. We assume, for example, that the plaque is initially devoid of SMCs until they enter the plaque as cap-forming cells \cite{misra2018integrin}. This particular assumption is not consistent with human atherosclerosis because the human intima contains a resident SMC population that is present prior to plaque growth. Despite this difference, experimental studies have shown consistency of human pathology with mouse models in many aspects of plaque SMC behaviour, including in phenotypic switching to SDMs \cite{misra2018integrin,shankman2015klf4}. Thus, we anticipate that our findings should at least be qualitatively conserved in the case of human atherosclerosis, even if not all SMCs that differentiate into SDMs contribute to depletion of the fibrous cap.          
%In this paper we have assumed that SMCs which switch to a macrophage phenotype have come out of the fibrous cap which separates the inside of the plaque from the bloodstream. However, SMCs are also present in humans both in the intima and in the media (the layer of the artery wall immediately outside the intima in both mice and humans). These intimal and medial SMCs may also switch to a macrophage-like phenotype. Even in these lower layers of the vessel wall, a loss of SMCs will result in a loss of cells that can produce collagen and exercise other stabilising effects on the plaque. Hence, although our arguments are couched in terms of cap thickness, the argument that loss of SMCs increases plaque vulnerability remains substantially true if SMCs are lost from other parts of the plaque.

This model for vascular SMC phenotypic switching in atherosclerosis highlights the crucial role of SMC plasticity on the dynamics of cells and lipids during atherosclerotic plaque progression. The reported results offer potentially useful insights that may contribute to the development of therapeutic strategies aimed at stabilising plaques and mitigating adverse cardiovascular outcomes.

\newpage
\section*{Acknowledgments}
We acknowledge funding support (to MRM) from an Australian Research Council Discovery Grant (DP200102071).

\bibliographystyle{unsrt}
\end{document}